\documentclass[onecolumn]{revtex4-1}
\usepackage{mathtools}
\usepackage{graphicx}
\usepackage{subcaption}
\usepackage{dcolumn}
\usepackage{bm}
\usepackage{physics}
\usepackage{color}
\usepackage{amsfonts}
\usepackage[normalem]{ulem}
\usepackage{hyperref}
\usepackage[utf8]{inputenc}
\usepackage{float}
\DeclarePairedDelimiter\floor{\lfloor}{\rfloor}

\definecolor{red}{RGB}{199,13,29}

\begin{document}

\preprint{APS/123-QED}

\title{\textit{QBoost} for regression problems: solving partial differential equations}

\author{Caio B. D. Góes}
\affiliation{Departamento de F\' \i sica, Universidade Federal de Santa Catarina, Florian\'opolis, Brazil}

\author{Thiago O. Maciel}
\affiliation{Departamento de F\' \i sica, Universidade Federal de Santa Catarina, Florian\'opolis, Brazil}
\affiliation{Universidade Federal do Rio de Janeiro, Rio de Janeiro, Brazil}

\author{Eduardo I. Duzzioni} 

\affiliation{
 QuanBy Computa\c c\~ao Qu\^antica, Florianópolis, Santa Catarina, Brazil
}
\affiliation{Departamento de F\' \i sica, Universidade Federal de Santa Catarina, Florian\'opolis, Brazil}

\author{Giovani G. Pollachini}
\affiliation{
 QuanBy Computa\c c\~ao Qu\^antica, Florianópolis, Santa Catarina, Brazil
}%

\author{Rafael Cuenca}
\affiliation{Engenharia Aeroespacial, Universidade Federal de Santa Catarina, Joinville, Santa Catarina, Brazil}

\author{Juan P. L. C. Salazar}
\affiliation{Engenharia Aeroespacial, Universidade Federal de Santa Catarina, Joinville, Santa Catarina, Brazil}

\email{duzzioni@gmail.com}

\date{\today}

\begin{abstract}

A hybrid algorithm based on machine learning and quantum ensemble learning is proposed to find an approximate solution to a partial differential equation with good precision and favorable scaling in the required number of qubits. The classical part is composed by training several regressors (weak-learners), capable of solving a partial differential equation approximately using machine learning. The quantum part consists of adapting the \textit{QBoost} algorithm to solve regression problems to build an ensemble of classical learners. We have successfully applied our framework to solve the 1D Burgers' equation with viscosity, showing that the quantum ensemble method really improves the solutions produced by classical weak-learners. We also implemented the algorithm on the D-Wave Systems, confirming the best performance of the quantum solution compared to the simulated annealing and exact solver methods.

\end{abstract}

\maketitle


\section{Introduction}\label{sec:introduction}

Classical Machine Learning (ML) techniques have recently become an important tool for addressing problems in quantum mechanics and in the physical science \cite{Carleo2019}, with applications in solving the Schrödinger's equation \cite{Hermann2020} quantum tomography \cite{Kieferov2017, Torlai2018}, quantum control  \cite{August2017}, quantum phase transitions \cite{Canabarro2019b,Carrasquilla2017}, quantum chemistry \cite{Dral2020}, astronomical object recognition \citep{Hezaveh2017}, and validation of quantum experiments \cite{Agresti2019}.  In a nutshell, ML is able to solve difficult problems using complex models \cite{aaron} which otherwise are hard for human mind to conceive.

As quantum mechanics can benefit from classical ML techniques, ML can also be improved by quantum mechanics. We find several instances of research that translate the various classical ML models for quantum computing with the aim of obtaining some speedup in training \cite{Yoo2014, Cai2015} and data storage \cite{Yu2019,Pepper2019,Huang2020}. Some examples are quantum K-Nearest-Neighbor \cite{Dang2018, Wang2019}, quantum decision trees \cite{Farhi1998, Lu2013}, quantum generative adversarial network \cite{Lloyd2018, Zoufal2019}, and quantum kernel methods \cite{Schuld2019, Blank2020}.

The main idea of the ML model based on ensemble learning is to add several different models using weights to create a combined model that is better than all of its constituents individually \cite{Dietterich2000, Breiman2001}. Ensemble learning models have been used for problems of classification \cite{Rokach2009} and regression \cite{MendesMoreira2012} alike. The \textit{QBoost} \cite{Qboost1, Qboost2, Qboost3} algorithm was the pioneer in translating the ensemble learning model to the quantum realm by addressing classification problems. Subsequently, more ensemble learning models have been quantized \cite{Schuld2018, Abbas2020}.

In the field of predictions by regression, we can find several applications of ML algorithms for ``real world'' problems ranging from weather forecast \cite{SalcedoSanz2011, SalcedoSanz2020}, traffic flow \cite{Wu2015, Zhang2018}, solar radiation estimation \cite{Lou2016, Alizamir2020} to aerodynamic applications \cite{Zhang20202, Dupuis2018, Andrs2012, Richmond2020, Umetani2018}. In general, fluid mechanics problems rely on the solution of partial differential equations (PDEs). There are several classical ML methods in literature that propose to find the solution of these PDEs \cite{Sirignano2018, Samaniego2020, Ranade2021, Raissi2018, Regazzoni2019}.

In this work, we propose an adaptation of the \textit{QBoost} algorithm for regression problems and apply the algorithm to solve the 1D viscous Burgers' equation. The paper is organized as follows: in \S\ref{sec:qboost} we describe the adaptation of the QBoost algorithm to deal with regression problems; in \S\ref{sec:hibrido} we explain the hybrid classical-quantum algorithm used to find the solution of the PDE; in \S\ref{sec: resultado} we present and discuss the results obtained. The conclusions and future perspectives are given in \S\ref{sec:conclusion}.

\section{\label{sec:qboost} QBoost for regression problems}

Quantum annealing is an optimization procedure that exploits the phenomenon of quantum fluctuations and quantum tunneling to find the minimum value of an objective function probabilistically, since the system is at a non-zero temperature \cite{Albash2018, Kadowaki1998}. The basic principle of quantum annealing is grounded on the adiabatic theorem, which tells us that if 
the system starts at the ground state of a Hamiltonian, $H_{I}$, that is known and easy to prepare and then it is allowed to evolve adiabatically to a final Hamiltonian, $H_{F}$, it would remain in its eigenstate \cite{Albash2018}, containing the solution of the desired problem. In other words, the Hamiltonian $H(t)$ changes in time according to,
\begin{align}
    H(t)=A(t)H_{I}+B(t)H_{F},
\end{align}
where $A(t)$ and $B(t)$ define the annealing schedule and must satisfy the following restrictions: $A(0) \neq 0$, $B(0) = 0$ and $B(T) \neq 0$, $A (T)=0$, where $ T $ is the total evolution time \cite{Albash2018, Kadowaki1998}. As the objective function to be minimized is encoded in $ H_{F} $, after the annealing process we have the minimum of this function.

The adiabatic theorem \cite{kato1950adiabatic,messiah1962quantum}, which has been stated in many different ways (see \citet{Albash2018} for a review), tells us that for the occurrence of a transitionless state evolution, the total evolution time $T$ must satisfy \cite{sarandy2004consistency} ($\hbar=1$) 
\begin{align}
    T \gg \frac{\max\limits_{0 \le s \le 1} \abs{\langle k(s)|\frac{dH(s)}{ds}|m(s) \rangle}}{\left( \min\limits_{0 \le s \le 1}|\Delta_{mk}(s)| \right)^2} \, ,
\end{align}
where $ \Delta_{mk} = E_{m} -E_{k}$, $ E_{m} $ and $ E_{k} $ are the eigenenergies associated to the eigenstates $|m(s)\rangle$ and $|k(s)\rangle$, respectively, $s=t/T$ is a dimensionless time, and $t$ is the current time. 

The Ising Hamiltonian in a transverse field is typically used to perform the quantum annealing process~\cite{Kadowaki1998}, as is the case of D-Wave quantum process units (QPU) \cite{dwave_qpu_2021}, given by,
\begin{align}
\label{eq:Ising}
    H(t)=-\frac{A(t)}{2}\sum_{i}\sigma_{x}^{i}+\frac{B(t)}{2}\left ( \sum_{i}d_{i}\sigma_{z}^{i}+ \sum_{i>j}D_{ij}\sigma_{z}^{i}\sigma_{z}^{j}\right)\,,
\end{align}
where \noindent $ \sigma_{x}^i $ and $ \sigma_{z}^i $ are Pauli matrices acting on the $i$-th qubit in $x$ and $z$ directions, $ d_{i} $ is the transverse field applied to the $i$-th qubit and $ D_{ij} $ is the coupling constant between the $i$-th and $j$-th qubits. 

In this work, we propose a variant of the \textit{QBoost} algorithm \cite{Qboost1, Qboost2, Qboost3} adequate for regression problems. In \textit{QBoost}, the cost function present in the ensemble learning algorithms is mapped to an Ising-type Hamiltonian and the optimization is performed via quantum annealing. Here, we perform similar steps, but under different constraints. The basic model of ensemble learning consists in creating an ensemble of $ K $ learners $ h_{k}(x) $ combined in a weighted sum to perform better than each one solely. These learners are trained given a few layers and few neurons, so we call them \emph{weak-learners}. Thus, we can evaluate a real function $f(x)$ as
\begin{align}
    f(x)=  \sum_{k=1}^{K}w_k h_{k}(x)\,,
\end{align}
 with $ x, f(x) \in \mathbb{R} $, and assuming the learners were trained beforehand, such that the parameters of their models could be omitted. 
 
 Training an ensemble model means finding the associated weights $w_k$ for each learner. For this task, we usually minimize two terms simultaneously: a \textit{loss function} $L (w,h,x) $ and a \textit{regularization} $R (w)$. The loss function, chosen as a convex least-squares function, 
\begin{align}
\label{eq:loss}
    L (w,h,y)= \frac{1}{M} \sum_{m = 1}^{M} \left[ \sum_{k=1}^{K}w_k h_{k}(x^m)-y^{m}  \right]^{2},
\end{align}
estimates the mean squared error that any regressor candidate imposes in a set of $ M $ training examples $ \{ h(x^{m}, y^{m} | m = 1, ..., M) \}$ in relation to  the true values $y^m$ provided in this training set. The regularization, as described in \citet{Qboost1}, aims to control the overfitting, and a natural choice for $ R (w) $ is an $ l_{0}$-norm penalization of $w$, which takes the weights to zero, if possible. However, $ l_{0}$-norm regularization leads to a non-convex optimization problem, and we replaced it by $ l_{2}$-norm, so we have a convex and differentiable function. In this way we must find the weights $w$ such that,
\begin{align}
    K(w,h,y)=\arg \min_{w} \left\{ J(w,h,y)\right\}\,,
\end{align}
where the total loss function is defined by,
\begin{align}
\label{eq:loss_total}
    J(w,h(x),y)= L(w,h(x),y)+\lambda ||w||_{2}^{2}\,,
\end{align}
where $ ||.||_{2} $ is the regularization of $ l_{2}$-norm and $ \lambda $ is an empirical parameter that controls the strength of the regularization. The expression above can be written explicitly as,
\begin{align}
\label{eq:J_qboost}
    J(w,h,y)=\frac{1}{M}\sum_{k,k^{'}=1}^{K}w_{k}w_{k^{'}}\left( \sum_{m=1}^{M}h_{k}(x^{m})h_{k^{'}}(x^{m}) \right)+\sum_{k=1}^{K}w_{k}\left[  \lambda w_k -2h_{k}(x^{m})y^{m} \right]\,,
\end{align}
where the terms not proportional to weights $w_k$ were discarded, as they do not influence the minimization process. Eq. \Ref{eq:J_qboost} is already in the form of a quadratic unconstrained binary optimization (QUBO) problem \cite{glover2018tutorial}, or a classical Ising-like Hamiltonian (see Eq.~\ref{eq:Ising}). 



Since we a interested in a regression problem, we need to write our weights as a $R$-bits floating-point approximation of the real value $w_i$. Following the methodology described in \citet{Rogers2020}, we apply the floating-point expansion to represent the weights $w_k$ in Eq. (\Ref{eq:constrain}). For any number $ \chi \in [0,2) $, the binary representation with accuracy of $R$ bits of resolution can be expressed by a string of bits $ [Q_{0} Q_{1} Q_{2} \cdots Q_{R} ]_{2} $, where $ Q_{r} \in \{0,1 \} $ is the value of the $ r $-th bit, and the square bracket indicates the binary representation. In terms of a $ 2^{-r} $ power series, we have,
\begin{align}
\label{eq:chi}
    \chi_{k} = \sum_{r=0}^{R-1}2^{-r}Q_{r}^{k}.
\end{align}
In order to represent the weights in a less restrictive domain, $ w_k \in [-d, 2c -d) $, we scale and shift $ \chi_k $ by,
\begin{align}
\label{eq:reconstrução}
    w_k=c\chi_{k} -d. \hspace{20pt}
\end{align}
When $ d> 0 $ and $ c> d / 2 $, the $ w_k $ domain will always have a positive and negative region, and the precise value of $c$ and $d$ can be chosen according to the specific problem. 

For the regression problem, we also added an extra constraint, $ \sum_{k = 1}^{K} w_{k} = 1 $, to reduce one variable and force an affine mixture of the weights. Thus, replacing in Eq. \Ref{eq:J_qboost}, we have,
\begin{align}
\label{eq:constrain}
\begin{split}
    J(w,h,y)=&\sum_{k,k^{'}=1}^{K-1}w_{k}w_{k^{'}}\left [\frac{1}{M}\sum_{m=1}^{M}(h_{k}^{m}-h_{K}^{m})(h_{k^{'}}^{m}-h_{K}^{m}) + \lambda (1+\delta_{k,k^{'}}) \right] + \\
    &+2\sum_{k=1}^{K-1}w_{k}\left [ \frac{1}{M}\sum_{m=1}^{M}(h_{K}^{m}-y^{m})(h_{K}^{m}-h_{k}^{m})   \right].
\end{split}
\end{align}

Rewriting Eq. (\ref{eq:constrain}) using Eq. (\ref{eq:chi}) and the scaling transformation of Eq. (\ref{eq:reconstrução}), the cost function becomes, 
\begin{align}
\begin{split}\label{eq:J}
        &J(h, y) =\sum_{k,k^{'}=1}^{K-1}c^{2}\left[\frac{1}{M}\sum_{m=1}^{M}(h_{k}^{m}-h_{K}^{m})(h_{k^{'}}^{m}-h_{K}^{m})+\lambda (1+\delta_{k,k^{'}}) \right]\sum_{r=0}^{R-1}2^{-r}Q_{r}^{k}\sum_{r^{'}=0}^{R-1}2^{-r^{'}}Q_{r^{'}}^{k^{'}}\\
        &+\sum_{k=1}^{K-1}2c\left[\frac{1}{M}\sum_{m=1}^{M}(h_{k}^{m}-h_{K}^{m})\left((h_{K}^{m}-y^{m})-d\sum_{k^{'}=1}^{K-1}(h_{k^{'}}^{m}-h_{K}^{m})  \right) - \lambda (1-dK) \right]\sum_{r=0}^{R-1}2^{-r}Q_{r}^{k}.
\end{split}
\end{align}

The physical qubits in the D-Wave processor are accessed by a 1-dimensional linear index, so it is necessary to merge the $ (k, r) $ indices into a single index $ l = R,R+1, ..., KR-1 $ using,
\begin{align}
    l(k,r)=kR+r\,,
\end{align}
with the inverse map given by $ k_{l} = \floor{l/R}$ and $ r_{l} = l \mod R $. Now we proceed to the quantization of the variables $Q_l$, which satisfy the eigenvalue equation $\hat{Q_{l}}\ket{Q}=Q_{l}\ket{Q}$, where the idempotence condition is imposed $\hat{Q_{l}}^{2}=\hat{Q_{l}}$, implying that the eigenvalues $Q_{l}\in \{0,1\}$. The eigenvectors are represented by $\ket{Q}=\ket{Q_{R}}\otimes \ket{Q_{R+1}}\otimes...\otimes\ket{Q_{(KR-1)}}$. Therefore, the quantization of Eq. (\ref{eq:J}) gives rise to the Hamiltonian,
\begin{align}
\label{eq:final}
\begin{split}
        &H(Q, h, y)= \sum_{l=R, l\neq l^{'}}^{KR-1}c^{2}2^{-(r_{l}+r_{l^{'}})}\left [ \frac{1}{M}\sum_{m=1}^{M}(h_{k_{l}}^{m}-h_{K}^{m})(h_{k_{l^{'}}}^{m}- h_{K}^{m})+\lambda(1+\delta_{k_{l},k_{l^{'}}}) \right]Q_{l}Q_{l^{'}}+ \\
        &+\sum_{l=R}^{KR-1}c2^{-r_{l}+1}\left \{  \frac{1}{M}\sum_{m=1}^{M}(h_{k_{l}}^{m}-h_{K}^{m})\left[(h_{K}^{m}-y^{m})-d\sum_{k_{l^{'}}=1}^{K-1}(h_{k_{l^{'}}}^{m}-h_{K}^{m})+c2^{-r_{l}-1}(h_{k_{l}}^{m}-h_{K}^{m})  \right] -\lambda (1+dK +c2^{-r_{l}}) \right \}Q_{l}.
\end{split}
\end{align}
Comparing the above equation with the QUBO Hamiltonian given by,
\begin{align}
\label{eq:qubo_final}
     H_{QUBO} = \sum_{i} \alpha_i Q_i + \sum_{i\ne j} \beta_{ij} Q_iQ_j\,,  
\end{align}
we identify the coefficients of the final Hamiltonian that are the input parameters for quantum annealing,
\begin{align}
\label{eq:alpha}
    \alpha_{i}= c2^{-r_{i}+1}\left\{ \frac{1}{M}\sum_{m=1}^{M}(h_{k_{i}}^{m}-h_{K}^{m})\left[ (h_{K}^{m}-y^{m})-d\left(\sum_{k_{j}=1}^{K-1}(h_{k_{j}}^{m}-h_{K}^{m})\right) +c2^{-r_{i}-1}(h_{k_{i}}^{m}-h_{K}^{m}) \right] - \lambda(1+dK+c2^{-r_{i}}) \right\}\,,
\end{align}
and
\begin{align}
\label{eq:beta}
    \beta_{ij}=c^{2}2^{-(r_{i}+r_{j})}\left[ \frac{1}{M}\sum_{m=1}^{M}(h_{k_{i}}^{m}-h_{K}^{m})(h_{k_{j}}^{m}-h_{K}^{m}) + \lambda(1+\delta_{k_{i},k_{j}}) \right]. 
\end{align}

\section{\label{sec:hibrido} Hybrid algorithm}

In this work we propose a hybrid algorithm that is able to perform regression in PDEs solutions. The classical part of the algorithm consists in creating neural networks (NN) capable of approaching the PDE solution, while the quantum part will create a boosting algorithm from the ensemble of classical NN (the weak-learners) to generate a strong NN. The quantum step of the algorithm is performed through the quantum annealing, which obtains the closest solution to the ground state of the Hamiltonian (Eq. (\ref{eq:final})), or equivalently, it finds the weights of a stronger NN.

\subsection{Classical part}

To introduce the classical part of the algorithm, let us consider a general $d$-dimensional parabolic PDE \footnote{The development of the QBoost method applied to regression problems and application to solve this PDE, which recovers the Burgers's equation as particular case, was motivated by the Airbus Quantum Computing Challenge, www.air bus.com/qc-challenge.html.}:
\begin{align}
\label{eq:PBE}
\begin{split}
    &\frac{\partial}{\partial t}u(t,x)+\mathcal{L}u(t,x)=0; \hspace{0.5cm} (t,x) \in [0,T] \times \Omega, \\
    &u(0,x)=u_{0}(x),  \\
    &u(t,x)=g(t,x), \; x \in \partial \Omega.
\end{split}
\end{align}
where $x \in \Omega \subset \mathbb{R}^d$, $t$ is time, $\mathcal{L}$ is an operator containing spatial derivatives, $u(t,x)$ is the solution of the PDE, $u_0(x)$ is the initial condition, and $g(t,x)$ is the boundary condition. This part of the algorithm performs a regression in the PDE solution through a NN, approximating the solution $ u(t,x) \approx f(t,x, \theta)$, where $ \theta $ is a set of real variables used to represent the internal parameters of the NN. Following~\citet{Sirignano2018}, we use as a measure of approximation the following cost function $ J(f) $,
\begin{align}
\begin{split}
\label{eq:custoclassico}
    J(f)=\left|\left| \frac{\partial}{\partial t}f(t,x,\theta)+ \mathcal{L}f(t,x,\theta) \right|\right|^{2}_{[0,T]\times\Omega,\rho_{1}}
    + ||f(t,x,\theta)- g(t,x)||^{2}_{[0,T]\times\partial\Omega,\rho_{2}}+||f(0,x,\theta) -u_{0}(x)||^{2}_{\Omega,\rho_{3}}\,, 
\end{split}
\end{align}
where $ ||f(y)||^{2}_{\mathcal{Y}, \rho} = \int_{\mathcal{Y}} | f (y) |^{2} \rho(y) dy $ and $ \rho(y) $ is a probability density over the domain $ y \in \mathcal{Y} $. $ J(f) $ measures how well $ f(t,x,\theta) $ satisfies the operators of the differential equation, boundary conditions and initial conditions. Therefore, if $ J(f) \approx 0 $, then $ f (t, x, \theta) $ fits the solution of the PDE.

The NN role is to find the parameters $ \theta $ which minimize the cost function $ J(f) $. As $ J(f) \rightarrow 0 $ the NN solution approaches the PDE solution, i.e., $ f(t,x, \theta) \rightarrow u (t,x) $. The advantage of this algorithm when compared to standard approaches in computational fluid dynamics is that it is not necessary to create a mesh, which is computationally expensive.

As described in \citet{Sirignano2018}, the algorithm consists of the following steps:

\begin{enumerate}
    \item Generate random points $ (t_{n}, x_{n}) \in [0, T] \times \Omega $, $ (\tau_{n}, z_{n}) \in [ 0, T] \times \partial \Omega $ and draw random points $ a_{n} $ in the domain $ \Omega $ according to the respective probability densities $ \rho_{1} $, $ \rho_{2}, $ and $ \rho_{3} $.
    
    \item Calculate the quadratic error $ G (\theta_{n}, s_{n}) $ on the points drawn in 1. using the set $ s_{n} = \{(t_{n}, x_{n}), (\tau_{n}, z_{n}), a_{n} \}$
    \begin{align}
    \label{eq:quadraticoclassico}
        G (\theta_{n}, s_{n}) = \left (\pdv{f}{t} (t_{n}, x_{n}, \theta_{n}) + \mathcal{L} f ( t_{n}, x_{n}, \theta_{n}) \right)^{2} + \left (f (\tau_{n}, z_{n}, \theta_{n}) - g (\tau_{n}, z(n)) \right)^{2} + \left (f (0, a_{n}, \theta_{n}) - u_{0} (a_{n}) \right)^{2}.
    \end{align}
    
    \item Update the weights $ \theta_{n} $ using steepest gradient,
    \begin{align}
      \theta_{n + 1} = \theta_{n} - \alpha_{n} \nabla_{\theta} G(\theta_{n}, s_{n})\,,
    \end{align}
     where $ \alpha_{n} $ is called a learning rate and decreases as $ n $ increases.
    
    \item Repeat the steps until the convergence criterion is reached $G(\theta_{n},s_{n})\leq \epsilon$.
\end{enumerate}

The gradient $ \nabla_{\theta} G(\theta_{n}, s_{n}) $ is an unbiased estimate of $ \nabla_{\theta} J (f (\cdot, \theta_{n})) $,
\begin{align}
    \mathbb{E}[\nabla_{\theta}G(\theta_{n},s_{n})|\theta_{n}]=\nabla_{\theta}J(f(\cdot,\theta_{n})).
\end{align}
In this way, the stochastic descent gradient algorithm will on average take steps in the downward direction of the objective function $ J (f (\cdot, \theta)) $, that is, $ J (f (\cdot, \theta(n + 1))) <J (f (\cdot, \Theta_{n})) $ and so $ \theta_{n + 1} $ is a better parameter estimate than $ \theta_{n} $.

\subsection{\label{sec:quantum part} Quantum Part}

The quantum part of the algorithm consists of performing a boosting in the ensemble learning step through the quantum annealing method. Here we propose a variant of QBoost \cite{Qboost1, Qboost2, Qboost3}, more specifically, the work focuses on the study of a regression problem through ensemble in the form $f(t,x) = \sum_{k=1}^{K}w_k h_{k}(t,x) $, where $ (t, x) \in [0, T] \times \Omega $, $ h (t,x) \in \mathbb{R}^{K} $ is a vector of outputs from neural networks previously trained in $ (t, x) $, and $ \textbf{w} \in \mathbb{R}^{K} $ is a vector of weights to be optimized such that  $\sum_i w_i=1$.

We describe the algorithm as follows:

\begin{enumerate}
    \item Validate the weak-learners produced by the classical part of the algorithm to create the training and test sets with $ M $ and $ N $ samples, respectively.
    \item Calculate the QUBO coefficients with the training set using Eq. (\ref{eq:alpha}) and (\ref{eq:beta}). 
    \item Use the coefficients as input for D-Wave's quantum annealers.
    \item Read the binary string output from D-Wave Systems and use Eq. (\ref{eq:reconstrução}) to reconstruct the optimal weights.
    \item Construct the final output of the ensemble $f(t,x)$ with the rebuilt weights.
\end{enumerate}

\section{\label{sec: resultado} Results and Discussion}

In order to validate our approach we use a PDE that is well known and that has an analytical solution. The \textit{toy model} chosen was that of the Burgers' 1D equation with viscosity $\nu$~\citep{BURGERS1948171}. Considering Eq. (\ref{eq:PBE}) in the domain $ (t, x) \in [0, T] \times [0,2 \pi] $ with the following initial and boundary conditions.
\begin{align}
&  \frac{\partial u}{\partial t} + u \frac{\partial u}{\partial x} = \nu \frac{\partial^{2}u}{\partial x^{2}}, \\
&u(0,x)=-2\frac{\nu}{\phi(0,x)}\frac{d\phi}{dx}+4 \;, \; x \in [0,2\pi], \\
&u(t,0)=u(t,2\pi) \; , \; t \in [0,T].
\end{align}
Under these conditions, Burgers' equation presents an analytical solution in the form,
\begin{align}
    u(t,x)=-2\frac{\nu}{\phi(t,x)}\frac{d\phi}{dx}+4, \; \; (t,x) \in [0,T]\times [0,2\pi]\,,
\end{align}
where,
\begin{align}
    \phi(x,t)=\exp{\frac{-(x-4t)^{2}}{4\nu(t+1)}}+\exp{\frac{-(x+4t-2\pi)^{2}}{4\nu(t+1)}}.
\end{align}

In Tab. \ref{tab:weaklearners} we present the weak-learners corresponding to the classical part of the algorithm used to create the ensemble. All NN were trained $ 2,000$ instances through random drawing in the domain $ (t, x) \in  [0,0.5]\times [0,2\pi] $ to obtain $ M = 160,000 $ and $ N = 60,000 $ instances of training and testing, respectively. $ 20,000 $ points from the times $ t \in \{0.0,0.05,0.15,0.20,0.30,0.35,0.40,0.50 \} $ were used for the training set and $ 20,000 $ points from  $ t \in \{0.10,0.25,0.45 \} $ for the test set. The probability densities $\rho_1$, $\rho_2$, and $\rho_3$ were obtained considering the values of the functions $u(t,x)$, $u(0,x)$, and $u(t,0)=u(t,2\pi)$, respectively, evaluated for each pair $(t,x)$ as defined above, in which only the variable $x \in [0,2\pi]$ was random drawn. We chosen $\epsilon = 550$. 

\begin{table}[htpb]
\begin{ruledtabular}
\begin{tabular}{cccccc}
&weak-learner & Loss Function(test) & time/$\#$ epochs & ($\#$neurons) x ($\#$layers) & \\
\hline
&[120,30,120,30,120] & 0.004283 & 1.619587 &  2100 &\\
&[10,20,30]          & 0.001574 & 0.276350 &  180  &\\
&[20,20,20,20,20,20] & 0.002017 & 0.284775 &  720  &\\
&[60,30,10,10,30,60] & 0.008058 & 1.266518 &  760  &\\
\end{tabular}
\end{ruledtabular}
\caption{\label{tab:weaklearners}Weak-learners of the viscous 1D Burgers' equation. The weak-learner $ [l_1, l_2, \cdots, l_i, \cdots, l_q] $ represents a NN with $q$ layers containing $ l_i$ neurons in the i-\textit{th} layer. The values of the Loss function (\ref{eq:quadraticoclassico}) were calculated using the test set. The  time/$ \# $epochs parameter represents the difficulty of training an NN.}
\label{tab:avg-fidelity}
\end{table}

\begin{figure}
\begin{subfigure}{.5\textwidth}
  \centering
  \includegraphics[width=\columnwidth]{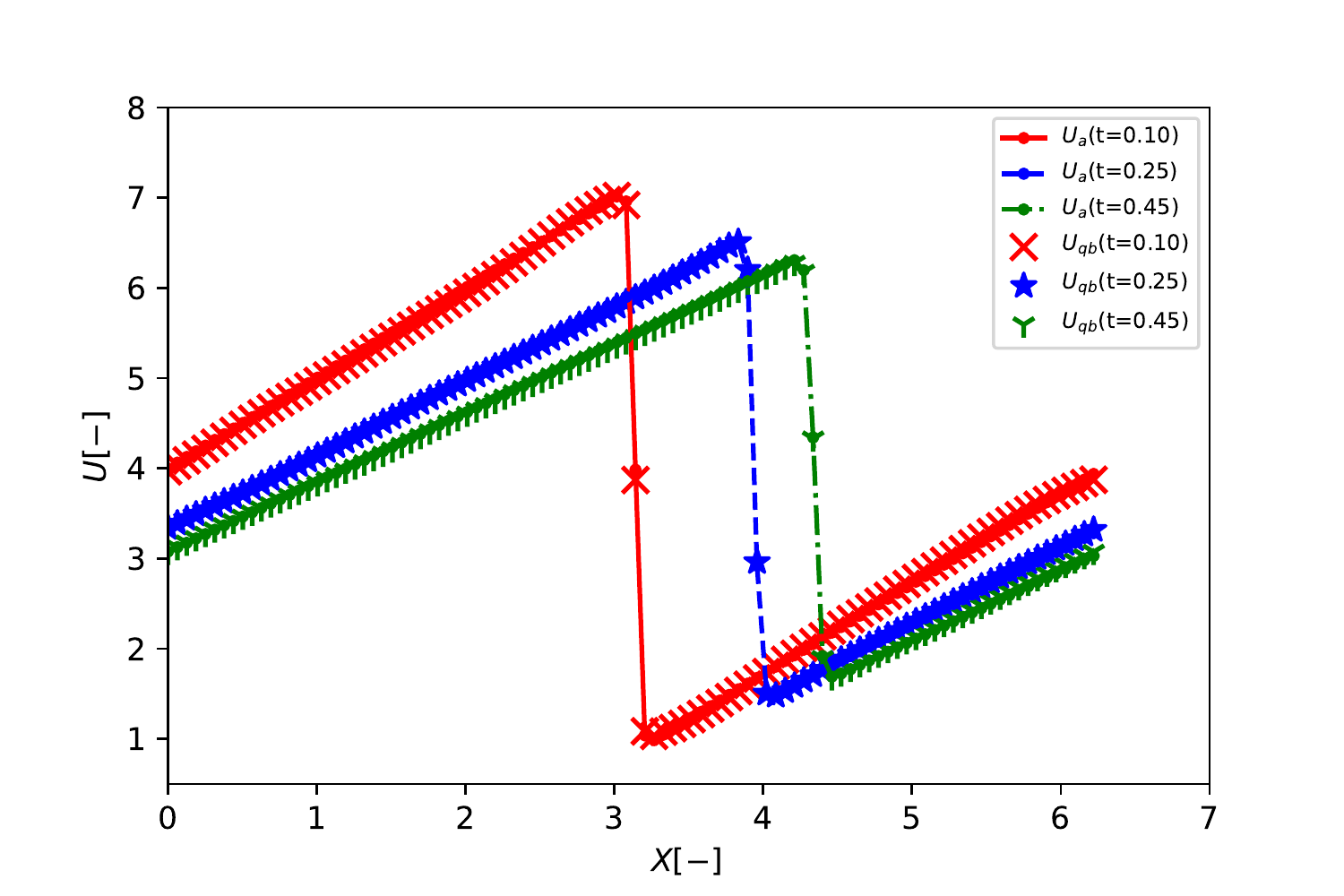}%
  \caption{Exact Solver.}
\end{subfigure}%
\begin{subfigure}{.5\textwidth}
  \centering
  \includegraphics[width=\columnwidth]{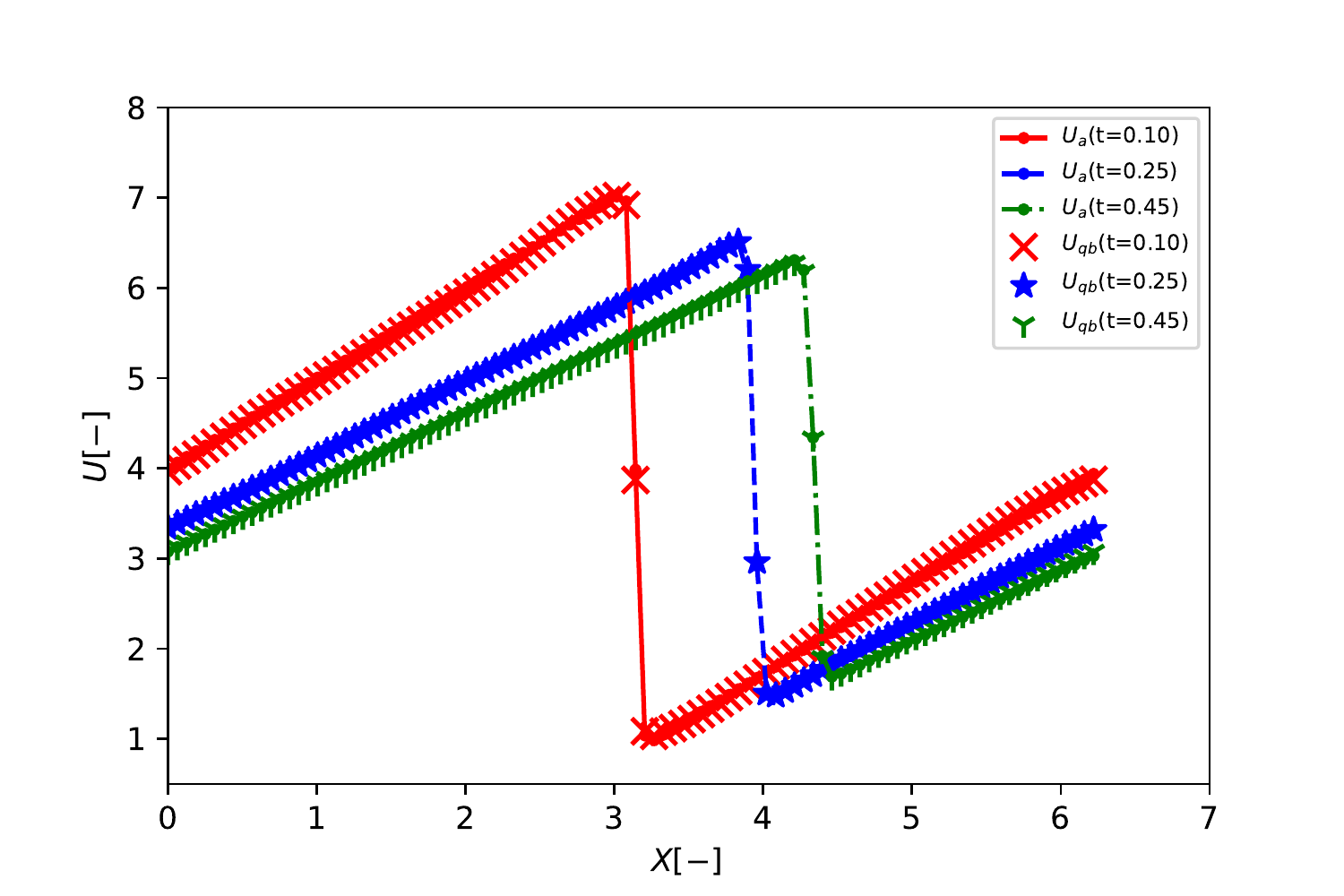}%
  \caption{Simulated Annealing.}
\end{subfigure}
\begin{subfigure}{.5\textwidth}
  \centering
  \includegraphics[width=\columnwidth]{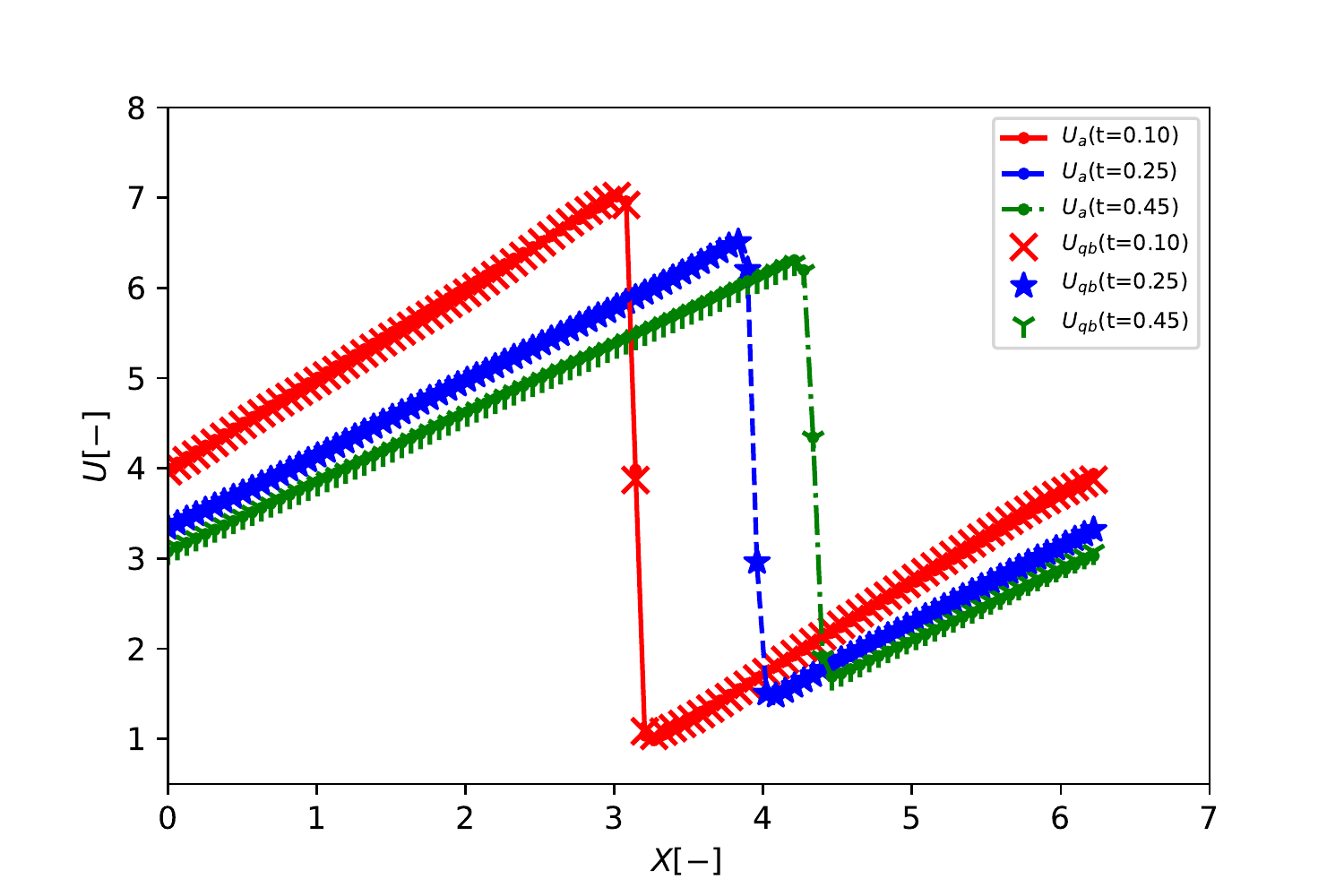}%
  \caption{Quantum Annealing.}
\end{subfigure}
\caption{\label{fig:ensemble} Analytical solution $ U_ {a}(x) $ and Qboost solution $ U_ {qb}(x) $ with $R=4$ of the test set for the 1D viscous Burgers' equation. We can see that both solutions are very similar in almost all extension of the graph. The solutions were obtained using three different functionalities of the D-Wave Ocean package: (a) exact diagonalization of Hamiltonian (\ref{eq:final}) through the Exact Solver, (b) classical solution through the simulated annealing, and (c) quantum solution through the 2000Q system.}
\end{figure}

We ran our algorithm solving the QUBO problem from Eq. (\ref{eq:qubo_final}) through the D-Wave Ocean Package in three different ways: $(a)$ using the Exact Solver function, $(b)$ the Simulating Annealing and, $(c)$ running in the 2000Q QPU. The weights of the ensemble are represented in the interval $w_k \in [-3,3]$, which means $c=d=3$. In Fig. \ref{fig:ensemble} we show the comparison between the solution produced by our ensemble, with $ \lambda = 0.0 $ (the criteria for choosing $\lambda$ can be seen in Appendix \ref{sec:B}), against the analytical solution of the $1D$ Burgers' equation for the times present in the test set. It is possible to observe that the solution produced by each method is very similar to the analytical solution. Although small, the difference between the solutions is more pronounced in the region where the function varies suddenly, i.e., the crosses and dots do not completely overlap.

\begin{figure}
\begin{subfigure}{.5\textwidth}
  \centering
  \includegraphics[width=\columnwidth]{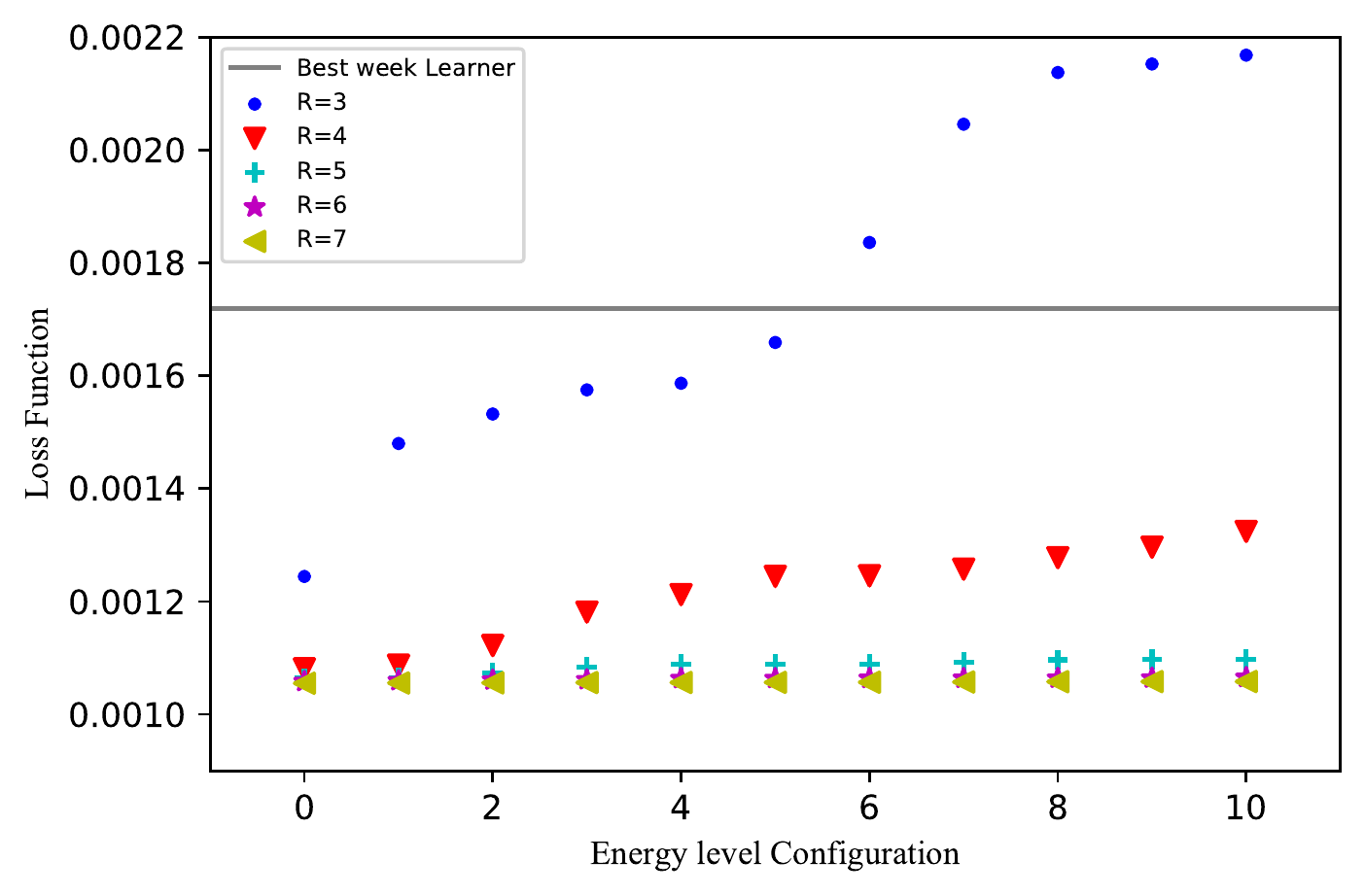}%
  \caption{Training set.}
\end{subfigure}%
\begin{subfigure}{.5\textwidth}
  \centering
  \includegraphics[width=\columnwidth]{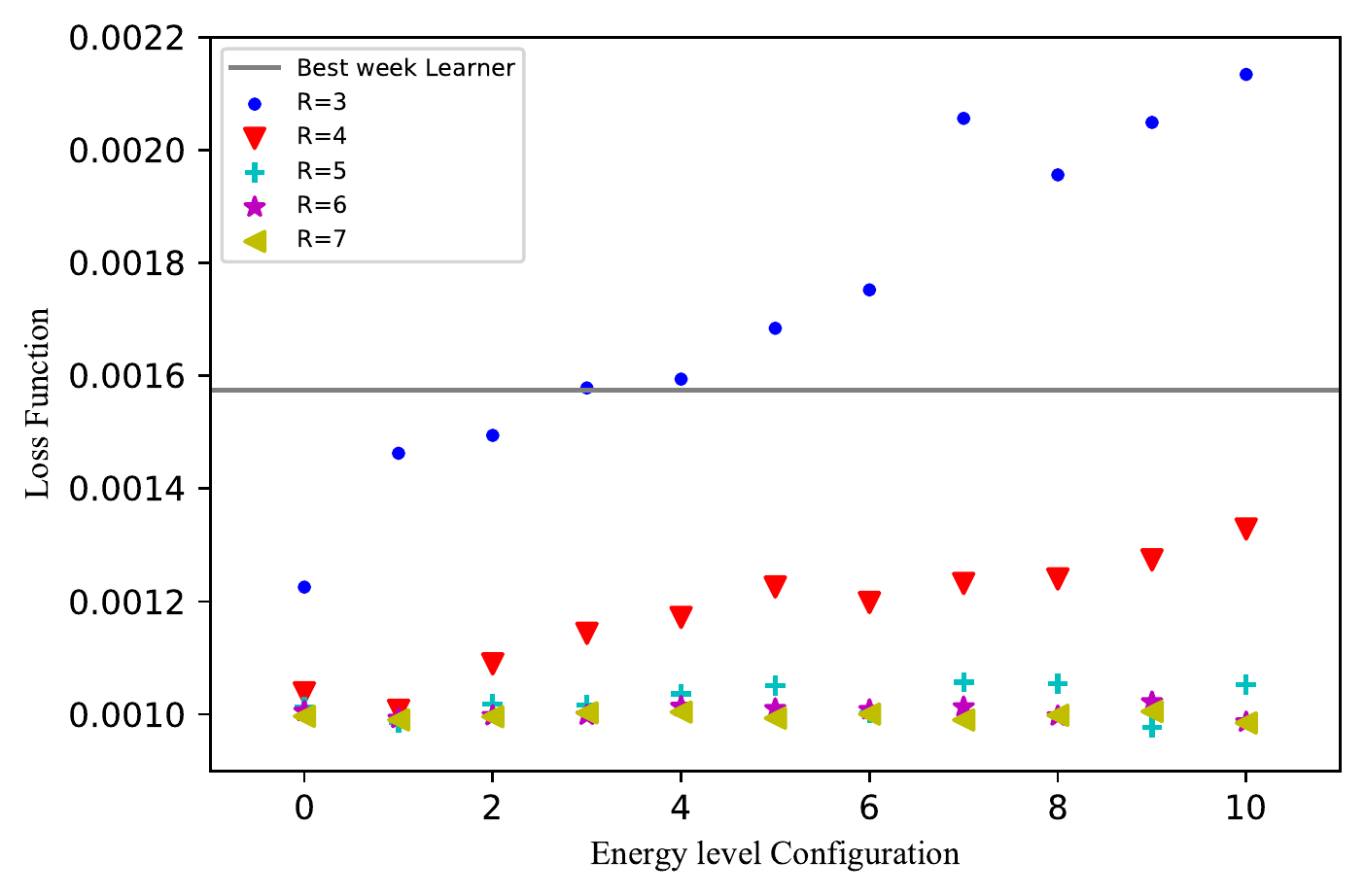}%
  \caption{Test set.}
\end{subfigure}
\caption{\label{fig:lossxenergy} Loss function values obtained from Qboost method for the first eleven energy levels of the final Hamiltonian (\ref{eq:final}) taking into account different numbers of precision qubits. The number $0$ represents the fundamental energy level of the final Hamiltonian. (a) and (b) refer to the training and test sets, respectively. In both graphs, all energy levels for precision $ R> 3 $ have an associated solution that is better than the one provided by the best weak-learner composing the ensemble, as shown by the horizontal line. } 
\end{figure}

The loss function (\ref{eq:final}) of our ensemble was evaluated in the test set, as shown in Table \ref{tab:precision} for different number of precision qubits $R$. In all cases the ensemble was able to perform better than all weak-learners used to compose them, see Table \ref{tab:weaklearners} and Fig. \ref{fig:lossxenergy}.  For the simulated annealing, we notice that the values of the loss function attain a minimum value of approximately $0.0010$, which is achieved with an accuracy of five qubits and is not reduced with the addition of more qubits. For quantum annealing we also observe a minimum in the values of the cost function around $0.0010$, which occurs for $R \ge 6$. 

We show the dependence of the loss function for different number of precision qubits $R = \{3,4,\cdots,14\}$ in Fig. \ref{fig:lossxenergy}, obtained from Qboost method for the (a) training and (b) test sets on the first $11$ energy levels of the final Hamiltonian (\ref{eq:final}). For sake of comparison, we also plot the loss function calculated of the best weak-learner (\emph{cf.} the horizontal line). As a result, the Qboost with $ R> 3 $  performs better than the best weak-learner for all energy levels displayed. We also observe that, as we increase the number of precision qubits, the values of the loss function become closer to each other for different energy levels of the final Hamiltonian. This is more pronounced in the training set (Fig. (\ref{fig:lossxenergy}a)). Indeed, this demonstrates that for a precision $R > 5$, even solutions that did not reach the ground state of the Hamiltonian can be considered valid, as the errors associated with these states are very close, within a predefined tolerance, $0.00005$. 
 
 The simulation using the Exact Solver has a limitation in relation to number of precision qubits due to the amount of memory required for its execution, which increases exponentially. We were able to achieve only $R=7$ in our personal computer of $16$ GB RAM. Note that, as we increase $R$, the computational time increased approximately by one order of magnitude. The time of the Simulated Annealing grows smoothly, compared to the Exact Solver, however, it starts from a value six orders of magnitude higher than the time spent by the Exact Solver. The quantum annealing time were fixed for all runs at $20 \mu s$. 
 

\begin{table}[htpb]
\begin{ruledtabular}
\begin{tabular}{ccccccccc}
&  &  Exact Solver &  &Simulated Annealing &  & Quantum Annealing & \\
&Precision (R) & Loss Function & Time (s) & Loss Function & Time (s) & Loss Function \\
\hline
&3 & 0.0012847& 0.002001 & 0.0012847 & 1111.0742 & 0.0012847 \\
&4 & 0.0010182& 0.016004 & 0.0010543 & 1654.2170 & 0.0010543\\
&5 & 0.0010131& 0.124011 & 0.0010131 & 2631.2131 & 0.0009863 \\
&6 & 0.0010052& 0.968069 & 0.0009856 & 4314.6367 & 0.0009927 \\
&7 & \textbf{0.0009967}& 10.027727& 0.0010084 & 4463.0229 & 0.0010227 \\
&8 & - & - & 0.0010191 &  5490.6962 & 0.0010250 \\
&9 & - & - & 0.0009908 &  6807.2573 & \textbf{0.0009779} \\
&10& - & - & 0.0010528 &  9473.4266 & 0.0010079 \\
&11& - & - & 0.0010324 & 11167.2495 & 0.0010291 \\
&12& - & - & \textbf{0.0009820} & 13716.8256 & 0.0009871 \\
&13& - & - & 0.0010566 & 14157.8434 & 0.0010012 \\
&14& - & - & 0.0009937 & 16095.1590 & 0.0010071 \\
\end{tabular}
\end{ruledtabular}
\caption{Values of Loss Function Eq. \ref{eq:loss_total} and the time needed to find the solution for different precision qubits in Eq. \ref{eq:final} and different methods. The parameter R is the precision of the floating-point expansion, while the methods used are special functions of the D-Wave Ocean package, providing the exact diagonalization through the Exact Solver, the classical annealing through the Simulated Annealing, and the quantum annealing ran in the 2000Q QPU. The annealing time is the same for all levels of precision in the quantum annealing solutions.}
\label{tab:precision}
\end{table}


\section{\label{sec:conclusion} Conclusion}

In this work, we propose an adaptation of the \textit{QBoost} algorithm to solve the regression problem using floating-point approximations to represent real variables. We applied it to solve a PDE, more precisely, the 1D Burgers' equation with viscosity. Our framework was applied successfully without the presence of overfitting and underfitting and the solution is in very good agreement with the analytical one (see Fig. \ref{fig:ensemble}).

We also conclude that the floating-point approximation should be used carefully, since the energy gap decreases rapidly increasing precision, although the accepted solutions are very close to the ground state, and therefore, are good approximations to the optimal solution. Another issue pertains to connectivity between qubits, which may limit the number of weak-learners used in the ensemble.

\begin{acknowledgments}
We acknowledge the financial support by Brazilian agencies CAPES, CNPq, and INCT-IQ (National Institute of Science and Technology for Quantum Information).
\end{acknowledgments}

\bibliography{main.bib}

\begin{thebibliography}{59}%
\makeatletter
\providecommand \@ifxundefined [1]{%
 \@ifx{#1\undefined}
}%
\providecommand \@ifnum [1]{%
 \ifnum #1\expandafter \@firstoftwo
 \else \expandafter \@secondoftwo
 \fi
}%
\providecommand \@ifx [1]{%
 \ifx #1\expandafter \@firstoftwo
 \else \expandafter \@secondoftwo
 \fi
}%
\providecommand \natexlab [1]{#1}%
\providecommand \enquote  [1]{``#1''}%
\providecommand \bibnamefont  [1]{#1}%
\providecommand \bibfnamefont [1]{#1}%
\providecommand \citenamefont [1]{#1}%
\providecommand \href@noop [0]{\@secondoftwo}%
\providecommand \href [0]{\begingroup \@sanitize@url \@href}%
\providecommand \@href[1]{\@@startlink{#1}\@@href}%
\providecommand \@@href[1]{\endgroup#1\@@endlink}%
\providecommand \@sanitize@url [0]{\catcode `\\12\catcode `\$12\catcode
  `\&12\catcode `\#12\catcode `\^12\catcode `\_12\catcode `\%12\relax}%
\providecommand \@@startlink[1]{}%
\providecommand \@@endlink[0]{}%
\providecommand \url  [0]{\begingroup\@sanitize@url \@url }%
\providecommand \@url [1]{\endgroup\@href {#1}{\urlprefix }}%
\providecommand \urlprefix  [0]{URL }%
\providecommand \Eprint [0]{\href }%
\providecommand \doibase [0]{http://dx.doi.org/}%
\providecommand \selectlanguage [0]{\@gobble}%
\providecommand \bibinfo  [0]{\@secondoftwo}%
\providecommand \bibfield  [0]{\@secondoftwo}%
\providecommand \translation [1]{[#1]}%
\providecommand \BibitemOpen [0]{}%
\providecommand \bibitemStop [0]{}%
\providecommand \bibitemNoStop [0]{.\EOS\space}%
\providecommand \EOS [0]{\spacefactor3000\relax}%
\providecommand \BibitemShut  [1]{\csname bibitem#1\endcsname}%
\let\auto@bib@innerbib\@empty
\bibitem [{\citenamefont {Carleo}\ \emph {et~al.}(2019)\citenamefont {Carleo},
  \citenamefont {Cirac}, \citenamefont {Cranmer}, \citenamefont {Daudet},
  \citenamefont {Schuld}, \citenamefont {Tishby}, \citenamefont
  {Vogt-Maranto},\ and\ \citenamefont {Zdeborov{\'{a}}}}]{Carleo2019}%
  \BibitemOpen
  \bibfield  {author} {\bibinfo {author} {\bibfnamefont {G.}~\bibnamefont
  {Carleo}}, \bibinfo {author} {\bibfnamefont {I.}~\bibnamefont {Cirac}},
  \bibinfo {author} {\bibfnamefont {K.}~\bibnamefont {Cranmer}}, \bibinfo
  {author} {\bibfnamefont {L.}~\bibnamefont {Daudet}}, \bibinfo {author}
  {\bibfnamefont {M.}~\bibnamefont {Schuld}}, \bibinfo {author} {\bibfnamefont
  {N.}~\bibnamefont {Tishby}}, \bibinfo {author} {\bibfnamefont
  {L.}~\bibnamefont {Vogt-Maranto}}, \ and\ \bibinfo {author} {\bibfnamefont
  {L.}~\bibnamefont {Zdeborov{\'{a}}}},\ }\href {\doibase
  10.1103/revmodphys.91.045002} {\bibfield  {journal} {\bibinfo  {journal}
  {Reviews of Modern Physics}\ }\textbf {\bibinfo {volume} {91}} (\bibinfo
  {year} {2019}),\ 10.1103/revmodphys.91.045002}\BibitemShut {NoStop}%
\bibitem [{\citenamefont {Hermann}\ \emph {et~al.}(2020)\citenamefont
  {Hermann}, \citenamefont {Sch\"{a}tzle},\ and\ \citenamefont
  {No{\'{e}}}}]{Hermann2020}%
  \BibitemOpen
  \bibfield  {author} {\bibinfo {author} {\bibfnamefont {J.}~\bibnamefont
  {Hermann}}, \bibinfo {author} {\bibfnamefont {Z.}~\bibnamefont
  {Sch\"{a}tzle}}, \ and\ \bibinfo {author} {\bibfnamefont {F.}~\bibnamefont
  {No{\'{e}}}},\ }\href {\doibase 10.1038/s41557-020-0544-y} {\bibfield
  {journal} {\bibinfo  {journal} {Nature Chemistry}\ }\textbf {\bibinfo
  {volume} {12}},\ \bibinfo {pages} {891} (\bibinfo {year} {2020})}\BibitemShut
  {NoStop}%
\bibitem [{\citenamefont {Kieferov{\'{a}}}\ and\ \citenamefont
  {Wiebe}(2017)}]{Kieferov2017}%
  \BibitemOpen
  \bibfield  {author} {\bibinfo {author} {\bibfnamefont {M.}~\bibnamefont
  {Kieferov{\'{a}}}}\ and\ \bibinfo {author} {\bibfnamefont {N.}~\bibnamefont
  {Wiebe}},\ }\href {\doibase 10.1103/physreva.96.062327} {\bibfield  {journal}
  {\bibinfo  {journal} {Physical Review A}\ }\textbf {\bibinfo {volume} {96}}
  (\bibinfo {year} {2017}),\ 10.1103/physreva.96.062327}\BibitemShut {NoStop}%
\bibitem [{\citenamefont {Torlai}\ \emph {et~al.}(2018)\citenamefont {Torlai},
  \citenamefont {Mazzola}, \citenamefont {Carrasquilla}, \citenamefont
  {Troyer}, \citenamefont {Melko},\ and\ \citenamefont {Carleo}}]{Torlai2018}%
  \BibitemOpen
  \bibfield  {author} {\bibinfo {author} {\bibfnamefont {G.}~\bibnamefont
  {Torlai}}, \bibinfo {author} {\bibfnamefont {G.}~\bibnamefont {Mazzola}},
  \bibinfo {author} {\bibfnamefont {J.}~\bibnamefont {Carrasquilla}}, \bibinfo
  {author} {\bibfnamefont {M.}~\bibnamefont {Troyer}}, \bibinfo {author}
  {\bibfnamefont {R.}~\bibnamefont {Melko}}, \ and\ \bibinfo {author}
  {\bibfnamefont {G.}~\bibnamefont {Carleo}},\ }\href {\doibase
  10.1038/s41567-018-0048-5} {\bibfield  {journal} {\bibinfo  {journal} {Nature
  Physics}\ }\textbf {\bibinfo {volume} {14}},\ \bibinfo {pages} {447}
  (\bibinfo {year} {2018})}\BibitemShut {NoStop}%
\bibitem [{\citenamefont {August}\ and\ \citenamefont {Ni}(2017)}]{August2017}%
  \BibitemOpen
  \bibfield  {author} {\bibinfo {author} {\bibfnamefont {M.}~\bibnamefont
  {August}}\ and\ \bibinfo {author} {\bibfnamefont {X.}~\bibnamefont {Ni}},\
  }\href {\doibase 10.1103/physreva.95.012335} {\bibfield  {journal} {\bibinfo
  {journal} {Physical Review A}\ }\textbf {\bibinfo {volume} {95}} (\bibinfo
  {year} {2017}),\ 10.1103/physreva.95.012335}\BibitemShut {NoStop}%
\bibitem [{\citenamefont {Canabarro}\ \emph {et~al.}(2019)\citenamefont
  {Canabarro}, \citenamefont {Fanchini}, \citenamefont {Malvezzi},
  \citenamefont {Pereira},\ and\ \citenamefont {Chaves}}]{Canabarro2019b}%
  \BibitemOpen
  \bibfield  {author} {\bibinfo {author} {\bibfnamefont {A.}~\bibnamefont
  {Canabarro}}, \bibinfo {author} {\bibfnamefont {F.~F.}\ \bibnamefont
  {Fanchini}}, \bibinfo {author} {\bibfnamefont {A.~L.}\ \bibnamefont
  {Malvezzi}}, \bibinfo {author} {\bibfnamefont {R.}~\bibnamefont {Pereira}}, \
  and\ \bibinfo {author} {\bibfnamefont {R.}~\bibnamefont {Chaves}},\ }\href
  {\doibase 10.1103/PhysRevB.100.045129} {\bibfield  {journal} {\bibinfo
  {journal} {Phys. Rev. B}\ }\textbf {\bibinfo {volume} {100}},\ \bibinfo
  {pages} {045129} (\bibinfo {year} {2019})}\BibitemShut {NoStop}%
\bibitem [{\citenamefont {Carrasquilla}\ and\ \citenamefont
  {Melko}(2017)}]{Carrasquilla2017}%
  \BibitemOpen
  \bibfield  {author} {\bibinfo {author} {\bibfnamefont {J.}~\bibnamefont
  {Carrasquilla}}\ and\ \bibinfo {author} {\bibfnamefont {R.~G.}\ \bibnamefont
  {Melko}},\ }\href {\doibase 10.1038/nphys4035} {\bibfield  {journal}
  {\bibinfo  {journal} {Nature Physics}\ }\textbf {\bibinfo {volume} {13}},\
  \bibinfo {pages} {431} (\bibinfo {year} {2017})}\BibitemShut {NoStop}%
\bibitem [{\citenamefont {Dral}(2020)}]{Dral2020}%
  \BibitemOpen
  \bibfield  {author} {\bibinfo {author} {\bibfnamefont {P.~O.}\ \bibnamefont
  {Dral}},\ }\href {\doibase 10.1021/acs.jpclett.9b03664} {\bibfield  {journal}
  {\bibinfo  {journal} {The Journal of Physical Chemistry Letters}\ }\textbf
  {\bibinfo {volume} {11}},\ \bibinfo {pages} {2336} (\bibinfo {year}
  {2020})}\BibitemShut {NoStop}%
\bibitem [{\citenamefont {Hezaveh}\ \emph {et~al.}(2017)\citenamefont
  {Hezaveh}, \citenamefont {Levasseur},\ and\ \citenamefont
  {Marshall}}]{Hezaveh2017}%
  \BibitemOpen
  \bibfield  {author} {\bibinfo {author} {\bibfnamefont {Y.~D.}\ \bibnamefont
  {Hezaveh}}, \bibinfo {author} {\bibfnamefont {L.~P.}\ \bibnamefont
  {Levasseur}}, \ and\ \bibinfo {author} {\bibfnamefont {P.~J.}\ \bibnamefont
  {Marshall}},\ }\href {\doibase 10.1038/nature23463} {\bibfield  {journal}
  {\bibinfo  {journal} {Nature}\ }\textbf {\bibinfo {volume} {548}},\ \bibinfo
  {pages} {555} (\bibinfo {year} {2017})}\BibitemShut {NoStop}%
\bibitem [{\citenamefont {Agresti}\ \emph {et~al.}(2019)\citenamefont
  {Agresti}, \citenamefont {Viggianiello}, \citenamefont {Flamini},
  \citenamefont {Spagnolo}, \citenamefont {Crespi}, \citenamefont {Osellame},
  \citenamefont {Wiebe},\ and\ \citenamefont {Sciarrino}}]{Agresti2019}%
  \BibitemOpen
  \bibfield  {author} {\bibinfo {author} {\bibfnamefont {I.}~\bibnamefont
  {Agresti}}, \bibinfo {author} {\bibfnamefont {N.}~\bibnamefont
  {Viggianiello}}, \bibinfo {author} {\bibfnamefont {F.}~\bibnamefont
  {Flamini}}, \bibinfo {author} {\bibfnamefont {N.}~\bibnamefont {Spagnolo}},
  \bibinfo {author} {\bibfnamefont {A.}~\bibnamefont {Crespi}}, \bibinfo
  {author} {\bibfnamefont {R.}~\bibnamefont {Osellame}}, \bibinfo {author}
  {\bibfnamefont {N.}~\bibnamefont {Wiebe}}, \ and\ \bibinfo {author}
  {\bibfnamefont {F.}~\bibnamefont {Sciarrino}},\ }\href {\doibase
  10.1103/physrevx.9.011013} {\bibfield  {journal} {\bibinfo  {journal}
  {Physical Review X}\ }\textbf {\bibinfo {volume} {9}} (\bibinfo {year}
  {2019}),\ 10.1103/physrevx.9.011013}\BibitemShut {NoStop}%
\bibitem [{\citenamefont {Goodfellow}\ \emph {et~al.}(2016)\citenamefont
  {Goodfellow}, \citenamefont {Bengio},\ and\ \citenamefont
  {Courville}}]{aaron}%
  \BibitemOpen
  \bibfield  {author} {\bibinfo {author} {\bibfnamefont {I.}~\bibnamefont
  {Goodfellow}}, \bibinfo {author} {\bibfnamefont {Y.}~\bibnamefont {Bengio}},
  \ and\ \bibinfo {author} {\bibfnamefont {A.}~\bibnamefont {Courville}},\
  }\href@noop {} {\emph {\bibinfo {title} {Deep Learning}}}\ (\bibinfo
  {publisher} {MIT Press},\ \bibinfo {year} {2016})\ \bibinfo {note}
  {\url{http://www.deeplearningbook.org}}\BibitemShut {NoStop}%
\bibitem [{\citenamefont {Yoo}\ \emph {et~al.}(2014)\citenamefont {Yoo},
  \citenamefont {Bang}, \citenamefont {Lee},\ and\ \citenamefont
  {Lee}}]{Yoo2014}%
  \BibitemOpen
  \bibfield  {author} {\bibinfo {author} {\bibfnamefont {S.}~\bibnamefont
  {Yoo}}, \bibinfo {author} {\bibfnamefont {J.}~\bibnamefont {Bang}}, \bibinfo
  {author} {\bibfnamefont {C.}~\bibnamefont {Lee}}, \ and\ \bibinfo {author}
  {\bibfnamefont {J.}~\bibnamefont {Lee}},\ }\href {\doibase
  10.1088/1367-2630/16/10/103014} {\bibfield  {journal} {\bibinfo  {journal}
  {New Journal of Physics}\ }\textbf {\bibinfo {volume} {16}},\ \bibinfo
  {pages} {103014} (\bibinfo {year} {2014})}\BibitemShut {NoStop}%
\bibitem [{\citenamefont {Cai}\ \emph {et~al.}(2015)\citenamefont {Cai},
  \citenamefont {Wu}, \citenamefont {Su}, \citenamefont {Chen}, \citenamefont
  {Wang}, \citenamefont {Li}, \citenamefont {Liu}, \citenamefont {Lu},\ and\
  \citenamefont {Pan}}]{Cai2015}%
  \BibitemOpen
  \bibfield  {author} {\bibinfo {author} {\bibfnamefont {X.-D.}\ \bibnamefont
  {Cai}}, \bibinfo {author} {\bibfnamefont {D.}~\bibnamefont {Wu}}, \bibinfo
  {author} {\bibfnamefont {Z.-E.}\ \bibnamefont {Su}}, \bibinfo {author}
  {\bibfnamefont {M.-C.}\ \bibnamefont {Chen}}, \bibinfo {author}
  {\bibfnamefont {X.-L.}\ \bibnamefont {Wang}}, \bibinfo {author}
  {\bibfnamefont {L.}~\bibnamefont {Li}}, \bibinfo {author} {\bibfnamefont
  {N.-L.}\ \bibnamefont {Liu}}, \bibinfo {author} {\bibfnamefont {C.-Y.}\
  \bibnamefont {Lu}}, \ and\ \bibinfo {author} {\bibfnamefont {J.-W.}\
  \bibnamefont {Pan}},\ }\href {\doibase 10.1103/physrevlett.114.110504}
  {\bibfield  {journal} {\bibinfo  {journal} {Physical Review Letters}\
  }\textbf {\bibinfo {volume} {114}} (\bibinfo {year} {2015}),\
  10.1103/physrevlett.114.110504}\BibitemShut {NoStop}%
\bibitem [{\citenamefont {Yu}\ \emph {et~al.}(2019)\citenamefont {Yu},
  \citenamefont {Gao}, \citenamefont {Lin},\ and\ \citenamefont
  {Wang}}]{Yu2019}%
  \BibitemOpen
  \bibfield  {author} {\bibinfo {author} {\bibfnamefont {C.-H.}\ \bibnamefont
  {Yu}}, \bibinfo {author} {\bibfnamefont {F.}~\bibnamefont {Gao}}, \bibinfo
  {author} {\bibfnamefont {S.}~\bibnamefont {Lin}}, \ and\ \bibinfo {author}
  {\bibfnamefont {J.}~\bibnamefont {Wang}},\ }\href {\doibase
  10.1007/s11128-019-2364-9} {\bibfield  {journal} {\bibinfo  {journal}
  {Quantum Information Processing}\ }\textbf {\bibinfo {volume} {18}} (\bibinfo
  {year} {2019}),\ 10.1007/s11128-019-2364-9}\BibitemShut {NoStop}%
\bibitem [{\citenamefont {Pepper}\ \emph {et~al.}(2019)\citenamefont {Pepper},
  \citenamefont {Tischler},\ and\ \citenamefont {Pryde}}]{Pepper2019}%
  \BibitemOpen
  \bibfield  {author} {\bibinfo {author} {\bibfnamefont {A.}~\bibnamefont
  {Pepper}}, \bibinfo {author} {\bibfnamefont {N.}~\bibnamefont {Tischler}}, \
  and\ \bibinfo {author} {\bibfnamefont {G.~J.}\ \bibnamefont {Pryde}},\ }\href
  {\doibase 10.1103/physrevlett.122.060501} {\bibfield  {journal} {\bibinfo
  {journal} {Physical Review Letters}\ }\textbf {\bibinfo {volume} {122}}
  (\bibinfo {year} {2019}),\ 10.1103/physrevlett.122.060501}\BibitemShut
  {NoStop}%
\bibitem [{\citenamefont {Huang}\ \emph {et~al.}(2020)\citenamefont {Huang},
  \citenamefont {Ma}, \citenamefont {Yin}, \citenamefont {Tang}, \citenamefont
  {Dong}, \citenamefont {Chen}, \citenamefont {Xiang}, \citenamefont {Li},\
  and\ \citenamefont {Guo}}]{Huang2020}%
  \BibitemOpen
  \bibfield  {author} {\bibinfo {author} {\bibfnamefont {C.-J.}\ \bibnamefont
  {Huang}}, \bibinfo {author} {\bibfnamefont {H.}~\bibnamefont {Ma}}, \bibinfo
  {author} {\bibfnamefont {Q.}~\bibnamefont {Yin}}, \bibinfo {author}
  {\bibfnamefont {J.-F.}\ \bibnamefont {Tang}}, \bibinfo {author}
  {\bibfnamefont {D.}~\bibnamefont {Dong}}, \bibinfo {author} {\bibfnamefont
  {C.}~\bibnamefont {Chen}}, \bibinfo {author} {\bibfnamefont {G.-Y.}\
  \bibnamefont {Xiang}}, \bibinfo {author} {\bibfnamefont {C.-F.}\ \bibnamefont
  {Li}}, \ and\ \bibinfo {author} {\bibfnamefont {G.-C.}\ \bibnamefont {Guo}},\
  }\href {\doibase 10.1103/physreva.102.032412} {\bibfield  {journal} {\bibinfo
   {journal} {Physical Review A}\ }\textbf {\bibinfo {volume} {102}} (\bibinfo
  {year} {2020}),\ 10.1103/physreva.102.032412}\BibitemShut {NoStop}%
\bibitem [{\citenamefont {Dang}\ \emph {et~al.}(2018)\citenamefont {Dang},
  \citenamefont {Jiang}, \citenamefont {Hu}, \citenamefont {Ji},\ and\
  \citenamefont {Zhang}}]{Dang2018}%
  \BibitemOpen
  \bibfield  {author} {\bibinfo {author} {\bibfnamefont {Y.}~\bibnamefont
  {Dang}}, \bibinfo {author} {\bibfnamefont {N.}~\bibnamefont {Jiang}},
  \bibinfo {author} {\bibfnamefont {H.}~\bibnamefont {Hu}}, \bibinfo {author}
  {\bibfnamefont {Z.}~\bibnamefont {Ji}}, \ and\ \bibinfo {author}
  {\bibfnamefont {W.}~\bibnamefont {Zhang}},\ }\href {\doibase
  10.1007/s11128-018-2004-9} {\bibfield  {journal} {\bibinfo  {journal}
  {Quantum Information Processing}\ }\textbf {\bibinfo {volume} {17}} (\bibinfo
  {year} {2018}),\ 10.1007/s11128-018-2004-9}\BibitemShut {NoStop}%
\bibitem [{\citenamefont {Wang}\ \emph {et~al.}(2019)\citenamefont {Wang},
  \citenamefont {Wang}, \citenamefont {Li}, \citenamefont {Adu-Gyamfi},
  \citenamefont {Tian},\ and\ \citenamefont {Zhu}}]{Wang2019}%
  \BibitemOpen
  \bibfield  {author} {\bibinfo {author} {\bibfnamefont {Y.}~\bibnamefont
  {Wang}}, \bibinfo {author} {\bibfnamefont {R.}~\bibnamefont {Wang}}, \bibinfo
  {author} {\bibfnamefont {D.}~\bibnamefont {Li}}, \bibinfo {author}
  {\bibfnamefont {D.}~\bibnamefont {Adu-Gyamfi}}, \bibinfo {author}
  {\bibfnamefont {K.}~\bibnamefont {Tian}}, \ and\ \bibinfo {author}
  {\bibfnamefont {Y.}~\bibnamefont {Zhu}},\ }\href {\doibase
  10.1007/s10773-019-04124-5} {\bibfield  {journal} {\bibinfo  {journal}
  {International Journal of Theoretical Physics}\ }\textbf {\bibinfo {volume}
  {58}},\ \bibinfo {pages} {2331} (\bibinfo {year} {2019})}\BibitemShut
  {NoStop}%
\bibitem [{\citenamefont {Farhi}\ and\ \citenamefont
  {Gutmann}(1998)}]{Farhi1998}%
  \BibitemOpen
  \bibfield  {author} {\bibinfo {author} {\bibfnamefont {E.}~\bibnamefont
  {Farhi}}\ and\ \bibinfo {author} {\bibfnamefont {S.}~\bibnamefont
  {Gutmann}},\ }\href {\doibase 10.1103/physreva.58.915} {\bibfield  {journal}
  {\bibinfo  {journal} {Physical Review A}\ }\textbf {\bibinfo {volume} {58}},\
  \bibinfo {pages} {915} (\bibinfo {year} {1998})}\BibitemShut {NoStop}%
\bibitem [{\citenamefont {Lu}\ and\ \citenamefont {Braunstein}(2013)}]{Lu2013}%
  \BibitemOpen
  \bibfield  {author} {\bibinfo {author} {\bibfnamefont {S.}~\bibnamefont
  {Lu}}\ and\ \bibinfo {author} {\bibfnamefont {S.~L.}\ \bibnamefont
  {Braunstein}},\ }\href {\doibase 10.1007/s11128-013-0687-5} {\bibfield
  {journal} {\bibinfo  {journal} {Quantum Information Processing}\ }\textbf
  {\bibinfo {volume} {13}},\ \bibinfo {pages} {757} (\bibinfo {year}
  {2013})}\BibitemShut {NoStop}%
\bibitem [{\citenamefont {Lloyd}\ and\ \citenamefont
  {Weedbrook}(2018)}]{Lloyd2018}%
  \BibitemOpen
  \bibfield  {author} {\bibinfo {author} {\bibfnamefont {S.}~\bibnamefont
  {Lloyd}}\ and\ \bibinfo {author} {\bibfnamefont {C.}~\bibnamefont
  {Weedbrook}},\ }\href {\doibase 10.1103/physrevlett.121.040502} {\bibfield
  {journal} {\bibinfo  {journal} {Physical Review Letters}\ }\textbf {\bibinfo
  {volume} {121}} (\bibinfo {year} {2018}),\
  10.1103/physrevlett.121.040502}\BibitemShut {NoStop}%
\bibitem [{\citenamefont {Zoufal}\ \emph {et~al.}(2019)\citenamefont {Zoufal},
  \citenamefont {Lucchi},\ and\ \citenamefont {Woerner}}]{Zoufal2019}%
  \BibitemOpen
  \bibfield  {author} {\bibinfo {author} {\bibfnamefont {C.}~\bibnamefont
  {Zoufal}}, \bibinfo {author} {\bibfnamefont {A.}~\bibnamefont {Lucchi}}, \
  and\ \bibinfo {author} {\bibfnamefont {S.}~\bibnamefont {Woerner}},\ }\href
  {\doibase 10.1038/s41534-019-0223-2} {\bibfield  {journal} {\bibinfo
  {journal} {npj Quantum Information}\ }\textbf {\bibinfo {volume} {5}}
  (\bibinfo {year} {2019}),\ 10.1038/s41534-019-0223-2}\BibitemShut {NoStop}%
\bibitem [{\citenamefont {Schuld}\ and\ \citenamefont
  {Killoran}(2019)}]{Schuld2019}%
  \BibitemOpen
  \bibfield  {author} {\bibinfo {author} {\bibfnamefont {M.}~\bibnamefont
  {Schuld}}\ and\ \bibinfo {author} {\bibfnamefont {N.}~\bibnamefont
  {Killoran}},\ }\href {\doibase 10.1103/physrevlett.122.040504} {\bibfield
  {journal} {\bibinfo  {journal} {Physical Review Letters}\ }\textbf {\bibinfo
  {volume} {122}} (\bibinfo {year} {2019}),\
  10.1103/physrevlett.122.040504}\BibitemShut {NoStop}%
\bibitem [{\citenamefont {Blank}\ \emph {et~al.}(2020)\citenamefont {Blank},
  \citenamefont {Park}, \citenamefont {Rhee},\ and\ \citenamefont
  {Petruccione}}]{Blank2020}%
  \BibitemOpen
  \bibfield  {author} {\bibinfo {author} {\bibfnamefont {C.}~\bibnamefont
  {Blank}}, \bibinfo {author} {\bibfnamefont {D.~K.}\ \bibnamefont {Park}},
  \bibinfo {author} {\bibfnamefont {J.-K.~K.}\ \bibnamefont {Rhee}}, \ and\
  \bibinfo {author} {\bibfnamefont {F.}~\bibnamefont {Petruccione}},\ }\href
  {\doibase 10.1038/s41534-020-0272-6} {\bibfield  {journal} {\bibinfo
  {journal} {npj Quantum Information}\ }\textbf {\bibinfo {volume} {6}}
  (\bibinfo {year} {2020}),\ 10.1038/s41534-020-0272-6}\BibitemShut {NoStop}%
\bibitem [{\citenamefont {Dietterich}(2000)}]{Dietterich2000}%
  \BibitemOpen
  \bibfield  {author} {\bibinfo {author} {\bibfnamefont {T.~G.}\ \bibnamefont
  {Dietterich}},\ }in\ \href {\doibase 10.1007/3-540-45014-9_1} {\emph
  {\bibinfo {booktitle} {Multiple Classifier Systems}}}\ (\bibinfo  {publisher}
  {Springer Berlin Heidelberg},\ \bibinfo {year} {2000})\ pp.\ \bibinfo {pages}
  {1--15}\BibitemShut {NoStop}%
\bibitem [{\citenamefont {Breiman}(2001)}]{Breiman2001}%
  \BibitemOpen
  \bibfield  {author} {\bibinfo {author} {\bibfnamefont {L.}~\bibnamefont
  {Breiman}},\ }\href {\doibase 10.1023/a:1010933404324} {\bibfield  {journal}
  {\bibinfo  {journal} {Machine Learning}\ }\textbf {\bibinfo {volume} {45}},\
  \bibinfo {pages} {5} (\bibinfo {year} {2001})}\BibitemShut {NoStop}%
\bibitem [{\citenamefont {Rokach}(2009)}]{Rokach2009}%
  \BibitemOpen
  \bibfield  {author} {\bibinfo {author} {\bibfnamefont {L.}~\bibnamefont
  {Rokach}},\ }\href {\doibase 10.1007/s10462-009-9124-7} {\bibfield  {journal}
  {\bibinfo  {journal} {Artificial Intelligence Review}\ }\textbf {\bibinfo
  {volume} {33}},\ \bibinfo {pages} {1} (\bibinfo {year} {2009})}\BibitemShut
  {NoStop}%
\bibitem [{\citenamefont {Mendes-Moreira}\ \emph {et~al.}(2012)\citenamefont
  {Mendes-Moreira}, \citenamefont {Soares}, \citenamefont {Jorge},\ and\
  \citenamefont {Sousa}}]{MendesMoreira2012}%
  \BibitemOpen
  \bibfield  {author} {\bibinfo {author} {\bibfnamefont {J.}~\bibnamefont
  {Mendes-Moreira}}, \bibinfo {author} {\bibfnamefont {C.}~\bibnamefont
  {Soares}}, \bibinfo {author} {\bibfnamefont {A.~M.}\ \bibnamefont {Jorge}}, \
  and\ \bibinfo {author} {\bibfnamefont {J.~F.~D.}\ \bibnamefont {Sousa}},\
  }\href {\doibase 10.1145/2379776.2379786} {\bibfield  {journal} {\bibinfo
  {journal} {{ACM} Computing Surveys}\ }\textbf {\bibinfo {volume} {45}},\
  \bibinfo {pages} {1} (\bibinfo {year} {2012})}\BibitemShut {NoStop}%
\bibitem [{\citenamefont {Neven}\ \emph {et~al.}(2009)\citenamefont {Neven},
  \citenamefont {Denchev}, \citenamefont {Rose},\ and\ \citenamefont
  {Macready}}]{Qboost1}%
  \BibitemOpen
  \bibfield  {author} {\bibinfo {author} {\bibfnamefont {H.}~\bibnamefont
  {Neven}}, \bibinfo {author} {\bibfnamefont {V.~S.}\ \bibnamefont {Denchev}},
  \bibinfo {author} {\bibfnamefont {G.}~\bibnamefont {Rose}}, \ and\ \bibinfo
  {author} {\bibfnamefont {W.~G.}\ \bibnamefont {Macready}},\ }\href@noop {}
  {\enquote {\bibinfo {title} {Training a large scale classifier with the
  quantum adiabatic algorithm},}\ } (\bibinfo {year} {2009}),\ \Eprint
  {http://arxiv.org/abs/arXiv:0912.0779} {arXiv:0912.0779} \BibitemShut
  {NoStop}%
\bibitem [{\citenamefont {Neven}\ \emph {et~al.}(2012)\citenamefont {Neven},
  \citenamefont {Denchev}, \citenamefont {Rose},\ and\ \citenamefont
  {Macready}}]{Qboost2}%
  \BibitemOpen
  \bibfield  {author} {\bibinfo {author} {\bibfnamefont {H.}~\bibnamefont
  {Neven}}, \bibinfo {author} {\bibfnamefont {V.~S.}\ \bibnamefont {Denchev}},
  \bibinfo {author} {\bibfnamefont {G.}~\bibnamefont {Rose}}, \ and\ \bibinfo
  {author} {\bibfnamefont {W.~G.}\ \bibnamefont {Macready}},\ }in\ \href
  {http://proceedings.mlr.press/v25/neven12.html} {\emph {\bibinfo {booktitle}
  {Proceedings of the Asian Conference on Machine Learning}}},\ \bibinfo
  {series} {Proceedings of Machine Learning Research}, Vol.~\bibinfo {volume}
  {25},\ \bibinfo {editor} {edited by\ \bibinfo {editor} {\bibfnamefont
  {S.~C.~H.}\ \bibnamefont {Hoi}}\ and\ \bibinfo {editor} {\bibfnamefont
  {W.}~\bibnamefont {Buntine}}}\ (\bibinfo  {publisher} {PMLR},\ \bibinfo
  {address} {Singapore Management University, Singapore},\ \bibinfo {year}
  {2012})\ pp.\ \bibinfo {pages} {333--348}\BibitemShut {NoStop}%
\bibitem [{\citenamefont {Neven}\ \emph {et~al.}(2008)\citenamefont {Neven},
  \citenamefont {Denchev}, \citenamefont {Rose},\ and\ \citenamefont
  {Macready}}]{Qboost3}%
  \BibitemOpen
  \bibfield  {author} {\bibinfo {author} {\bibfnamefont {H.}~\bibnamefont
  {Neven}}, \bibinfo {author} {\bibfnamefont {V.~S.}\ \bibnamefont {Denchev}},
  \bibinfo {author} {\bibfnamefont {G.}~\bibnamefont {Rose}}, \ and\ \bibinfo
  {author} {\bibfnamefont {W.~G.}\ \bibnamefont {Macready}},\ }\href@noop {}
  {\enquote {\bibinfo {title} {Training a binary classifier with the quantum
  adiabatic algorithm},}\ } (\bibinfo {year} {2008}),\ \Eprint
  {http://arxiv.org/abs/arXiv:0811.0416} {arXiv:0811.0416} \BibitemShut
  {NoStop}%
\bibitem [{\citenamefont {Schuld}\ and\ \citenamefont
  {Petruccione}(2018)}]{Schuld2018}%
  \BibitemOpen
  \bibfield  {author} {\bibinfo {author} {\bibfnamefont {M.}~\bibnamefont
  {Schuld}}\ and\ \bibinfo {author} {\bibfnamefont {F.}~\bibnamefont
  {Petruccione}},\ }\href {\doibase 10.1038/s41598-018-20403-3} {\bibfield
  {journal} {\bibinfo  {journal} {Scientific Reports}\ }\textbf {\bibinfo
  {volume} {8}} (\bibinfo {year} {2018}),\
  10.1038/s41598-018-20403-3}\BibitemShut {NoStop}%
\bibitem [{\citenamefont {Abbas}\ \emph {et~al.}(2020)\citenamefont {Abbas},
  \citenamefont {Schuld},\ and\ \citenamefont {Petruccione}}]{Abbas2020}%
  \BibitemOpen
  \bibfield  {author} {\bibinfo {author} {\bibfnamefont {A.}~\bibnamefont
  {Abbas}}, \bibinfo {author} {\bibfnamefont {M.}~\bibnamefont {Schuld}}, \
  and\ \bibinfo {author} {\bibfnamefont {F.}~\bibnamefont {Petruccione}},\
  }\href {\doibase 10.1007/s42484-020-00018-6} {\bibfield  {journal} {\bibinfo
  {journal} {Quantum Machine Intelligence}\ }\textbf {\bibinfo {volume} {2}}
  (\bibinfo {year} {2020}),\ 10.1007/s42484-020-00018-6}\BibitemShut {NoStop}%
\bibitem [{\citenamefont {Salcedo-Sanz}\ \emph {et~al.}(2011)\citenamefont
  {Salcedo-Sanz}, \citenamefont {Ortiz-Garc{\i}{\textasciiacute}a},
  \citenamefont {P{\'{e}}rez-Bellido}, \citenamefont {Portilla-Figueras},\ and\
  \citenamefont {Prieto}}]{SalcedoSanz2011}%
  \BibitemOpen
  \bibfield  {author} {\bibinfo {author} {\bibfnamefont {S.}~\bibnamefont
  {Salcedo-Sanz}}, \bibinfo {author} {\bibfnamefont {E.~G.}\ \bibnamefont
  {Ortiz-Garc{\i}{\textasciiacute}a}}, \bibinfo {author} {\bibfnamefont
  {{\'{A}}.~M.}\ \bibnamefont {P{\'{e}}rez-Bellido}}, \bibinfo {author}
  {\bibfnamefont {A.}~\bibnamefont {Portilla-Figueras}}, \ and\ \bibinfo
  {author} {\bibfnamefont {L.}~\bibnamefont {Prieto}},\ }\href {\doibase
  10.1016/j.eswa.2010.09.067} {\bibfield  {journal} {\bibinfo  {journal}
  {Expert Systems with Applications}\ }\textbf {\bibinfo {volume} {38}},\
  \bibinfo {pages} {4052} (\bibinfo {year} {2011})}\BibitemShut {NoStop}%
\bibitem [{\citenamefont {Salcedo-Sanz}\ \emph {et~al.}(2020)\citenamefont
  {Salcedo-Sanz}, \citenamefont {Ghamisi}, \citenamefont {Piles}, \citenamefont
  {Werner}, \citenamefont {Cuadra}, \citenamefont {Moreno-Mart{\'{\i}}nez},
  \citenamefont {Izquierdo-Verdiguier}, \citenamefont
  {Mu{\~{n}}oz-Mar{\'{\i}}}, \citenamefont {Mosavi},\ and\ \citenamefont
  {Camps-Valls}}]{SalcedoSanz2020}%
  \BibitemOpen
  \bibfield  {author} {\bibinfo {author} {\bibfnamefont {S.}~\bibnamefont
  {Salcedo-Sanz}}, \bibinfo {author} {\bibfnamefont {P.}~\bibnamefont
  {Ghamisi}}, \bibinfo {author} {\bibfnamefont {M.}~\bibnamefont {Piles}},
  \bibinfo {author} {\bibfnamefont {M.}~\bibnamefont {Werner}}, \bibinfo
  {author} {\bibfnamefont {L.}~\bibnamefont {Cuadra}}, \bibinfo {author}
  {\bibfnamefont {A.}~\bibnamefont {Moreno-Mart{\'{\i}}nez}}, \bibinfo {author}
  {\bibfnamefont {E.}~\bibnamefont {Izquierdo-Verdiguier}}, \bibinfo {author}
  {\bibfnamefont {J.}~\bibnamefont {Mu{\~{n}}oz-Mar{\'{\i}}}}, \bibinfo
  {author} {\bibfnamefont {A.}~\bibnamefont {Mosavi}}, \ and\ \bibinfo {author}
  {\bibfnamefont {G.}~\bibnamefont {Camps-Valls}},\ }\href {\doibase
  10.1016/j.inffus.2020.07.004} {\bibfield  {journal} {\bibinfo  {journal}
  {Information Fusion}\ }\textbf {\bibinfo {volume} {63}},\ \bibinfo {pages}
  {256} (\bibinfo {year} {2020})}\BibitemShut {NoStop}%
\bibitem [{\citenamefont {Wu}\ \emph {et~al.}(2015)\citenamefont {Wu},
  \citenamefont {Tan}, \citenamefont {Peter}, \citenamefont {Shen},\ and\
  \citenamefont {Ran}}]{Wu2015}%
  \BibitemOpen
  \bibfield  {author} {\bibinfo {author} {\bibfnamefont {Y.}~\bibnamefont
  {Wu}}, \bibinfo {author} {\bibfnamefont {H.}~\bibnamefont {Tan}}, \bibinfo
  {author} {\bibfnamefont {J.}~\bibnamefont {Peter}}, \bibinfo {author}
  {\bibfnamefont {B.}~\bibnamefont {Shen}}, \ and\ \bibinfo {author}
  {\bibfnamefont {B.}~\bibnamefont {Ran}},\ }in\ \href {\doibase
  10.1061/9780784479292.051} {\emph {\bibinfo {booktitle} {{CICTP} 2015}}}\
  (\bibinfo  {publisher} {American Society of Civil Engineers},\ \bibinfo
  {year} {2015})\BibitemShut {NoStop}%
\bibitem [{\citenamefont {Zhang}\ \emph {et~al.}(2018)\citenamefont {Zhang},
  \citenamefont {Alharbe}, \citenamefont {Luo}, \citenamefont {Yao},\ and\
  \citenamefont {Li}}]{Zhang2018}%
  \BibitemOpen
  \bibfield  {author} {\bibinfo {author} {\bibfnamefont {L.}~\bibnamefont
  {Zhang}}, \bibinfo {author} {\bibfnamefont {N.~R.}\ \bibnamefont {Alharbe}},
  \bibinfo {author} {\bibfnamefont {G.}~\bibnamefont {Luo}}, \bibinfo {author}
  {\bibfnamefont {Z.}~\bibnamefont {Yao}}, \ and\ \bibinfo {author}
  {\bibfnamefont {Y.}~\bibnamefont {Li}},\ }\href {\doibase
  10.26599/tst.2018.9010045} {\bibfield  {journal} {\bibinfo  {journal}
  {Tsinghua Science and Technology}\ }\textbf {\bibinfo {volume} {23}},\
  \bibinfo {pages} {479} (\bibinfo {year} {2018})}\BibitemShut {NoStop}%
\bibitem [{\citenamefont {Lou}\ \emph {et~al.}(2016)\citenamefont {Lou},
  \citenamefont {Li}, \citenamefont {Lam},\ and\ \citenamefont
  {Chan}}]{Lou2016}%
  \BibitemOpen
  \bibfield  {author} {\bibinfo {author} {\bibfnamefont {S.}~\bibnamefont
  {Lou}}, \bibinfo {author} {\bibfnamefont {D.~H.}\ \bibnamefont {Li}},
  \bibinfo {author} {\bibfnamefont {J.~C.}\ \bibnamefont {Lam}}, \ and\
  \bibinfo {author} {\bibfnamefont {W.~W.}\ \bibnamefont {Chan}},\ }\href
  {\doibase 10.1016/j.apenergy.2016.08.093} {\bibfield  {journal} {\bibinfo
  {journal} {Applied Energy}\ }\textbf {\bibinfo {volume} {181}},\ \bibinfo
  {pages} {367} (\bibinfo {year} {2016})}\BibitemShut {NoStop}%
\bibitem [{\citenamefont {Alizamir}\ \emph {et~al.}(2020)\citenamefont
  {Alizamir}, \citenamefont {Kim}, \citenamefont {Kisi},\ and\ \citenamefont
  {Zounemat-Kermani}}]{Alizamir2020}%
  \BibitemOpen
  \bibfield  {author} {\bibinfo {author} {\bibfnamefont {M.}~\bibnamefont
  {Alizamir}}, \bibinfo {author} {\bibfnamefont {S.}~\bibnamefont {Kim}},
  \bibinfo {author} {\bibfnamefont {O.}~\bibnamefont {Kisi}}, \ and\ \bibinfo
  {author} {\bibfnamefont {M.}~\bibnamefont {Zounemat-Kermani}},\ }\href
  {\doibase 10.1016/j.energy.2020.117239} {\bibfield  {journal} {\bibinfo
  {journal} {Energy}\ }\textbf {\bibinfo {volume} {197}},\ \bibinfo {pages}
  {117239} (\bibinfo {year} {2020})}\BibitemShut {NoStop}%
\bibitem [{\citenamefont {Zhang}\ and\ \citenamefont
  {Zhao}(2020)}]{Zhang20202}%
  \BibitemOpen
  \bibfield  {author} {\bibinfo {author} {\bibfnamefont {J.}~\bibnamefont
  {Zhang}}\ and\ \bibinfo {author} {\bibfnamefont {X.}~\bibnamefont {Zhao}},\
  }\href {\doibase 10.2514/1.j059877} {\bibfield  {journal} {\bibinfo
  {journal} {{AIAA} Journal}\ ,\ \bibinfo {pages} {1}} (\bibinfo {year}
  {2020})}\BibitemShut {NoStop}%
\bibitem [{\citenamefont {Dupuis}\ \emph {et~al.}(2018)\citenamefont {Dupuis},
  \citenamefont {Jouhaud},\ and\ \citenamefont {Sagaut}}]{Dupuis2018}%
  \BibitemOpen
  \bibfield  {author} {\bibinfo {author} {\bibfnamefont {R.}~\bibnamefont
  {Dupuis}}, \bibinfo {author} {\bibfnamefont {J.-C.}\ \bibnamefont {Jouhaud}},
  \ and\ \bibinfo {author} {\bibfnamefont {P.}~\bibnamefont {Sagaut}},\ }\href
  {\doibase 10.2514/1.j056405} {\bibfield  {journal} {\bibinfo  {journal}
  {{AIAA} Journal}\ }\textbf {\bibinfo {volume} {56}},\ \bibinfo {pages} {3622}
  (\bibinfo {year} {2018})}\BibitemShut {NoStop}%
\bibitem [{\citenamefont {Andr{\'{e}}s}\ \emph {et~al.}(2012)\citenamefont
  {Andr{\'{e}}s}, \citenamefont {Salcedo-Sanz}, \citenamefont {Monge},\ and\
  \citenamefont {P{\'{e}}rez-Bellido}}]{Andrs2012}%
  \BibitemOpen
  \bibfield  {author} {\bibinfo {author} {\bibfnamefont {E.}~\bibnamefont
  {Andr{\'{e}}s}}, \bibinfo {author} {\bibfnamefont {S.}~\bibnamefont
  {Salcedo-Sanz}}, \bibinfo {author} {\bibfnamefont {F.}~\bibnamefont {Monge}},
  \ and\ \bibinfo {author} {\bibfnamefont {A.}~\bibnamefont
  {P{\'{e}}rez-Bellido}},\ }\href {\doibase 10.1016/j.eswa.2012.02.197}
  {\bibfield  {journal} {\bibinfo  {journal} {Expert Systems with
  Applications}\ }\textbf {\bibinfo {volume} {39}},\ \bibinfo {pages} {10700}
  (\bibinfo {year} {2012})}\BibitemShut {NoStop}%
\bibitem [{\citenamefont {Richmond}\ \emph {et~al.}(2020)\citenamefont
  {Richmond}, \citenamefont {Sobey}, \citenamefont {Pandit},\ and\
  \citenamefont {Kolios}}]{Richmond2020}%
  \BibitemOpen
  \bibfield  {author} {\bibinfo {author} {\bibfnamefont {M.}~\bibnamefont
  {Richmond}}, \bibinfo {author} {\bibfnamefont {A.}~\bibnamefont {Sobey}},
  \bibinfo {author} {\bibfnamefont {R.}~\bibnamefont {Pandit}}, \ and\ \bibinfo
  {author} {\bibfnamefont {A.}~\bibnamefont {Kolios}},\ }\href {\doibase
  10.1016/j.renene.2020.07.083} {\bibfield  {journal} {\bibinfo  {journal}
  {Renewable Energy}\ }\textbf {\bibinfo {volume} {161}},\ \bibinfo {pages}
  {650} (\bibinfo {year} {2020})}\BibitemShut {NoStop}%
\bibitem [{\citenamefont {Umetani}\ and\ \citenamefont
  {Bickel}(2018)}]{Umetani2018}%
  \BibitemOpen
  \bibfield  {author} {\bibinfo {author} {\bibfnamefont {N.}~\bibnamefont
  {Umetani}}\ and\ \bibinfo {author} {\bibfnamefont {B.}~\bibnamefont
  {Bickel}},\ }\href {\doibase 10.1145/3197517.3201325} {\bibfield  {journal}
  {\bibinfo  {journal} {{ACM} Transactions on Graphics}\ }\textbf {\bibinfo
  {volume} {37}},\ \bibinfo {pages} {1} (\bibinfo {year} {2018})}\BibitemShut
  {NoStop}%
\bibitem [{\citenamefont {Sirignano}\ and\ \citenamefont
  {Spiliopoulos}(2018)}]{Sirignano2018}%
  \BibitemOpen
  \bibfield  {author} {\bibinfo {author} {\bibfnamefont {J.}~\bibnamefont
  {Sirignano}}\ and\ \bibinfo {author} {\bibfnamefont {K.}~\bibnamefont
  {Spiliopoulos}},\ }\href {\doibase 10.1016/j.jcp.2018.08.029} {\bibfield
  {journal} {\bibinfo  {journal} {Journal of Computational Physics}\ }\textbf
  {\bibinfo {volume} {375}},\ \bibinfo {pages} {1339} (\bibinfo {year}
  {2018})}\BibitemShut {NoStop}%
\bibitem [{\citenamefont {Samaniego}\ \emph {et~al.}(2020)\citenamefont
  {Samaniego}, \citenamefont {Anitescu}, \citenamefont {Goswami}, \citenamefont
  {Nguyen-Thanh}, \citenamefont {Guo}, \citenamefont {Hamdia}, \citenamefont
  {Zhuang},\ and\ \citenamefont {Rabczuk}}]{Samaniego2020}%
  \BibitemOpen
  \bibfield  {author} {\bibinfo {author} {\bibfnamefont {E.}~\bibnamefont
  {Samaniego}}, \bibinfo {author} {\bibfnamefont {C.}~\bibnamefont {Anitescu}},
  \bibinfo {author} {\bibfnamefont {S.}~\bibnamefont {Goswami}}, \bibinfo
  {author} {\bibfnamefont {V.}~\bibnamefont {Nguyen-Thanh}}, \bibinfo {author}
  {\bibfnamefont {H.}~\bibnamefont {Guo}}, \bibinfo {author} {\bibfnamefont
  {K.}~\bibnamefont {Hamdia}}, \bibinfo {author} {\bibfnamefont
  {X.}~\bibnamefont {Zhuang}}, \ and\ \bibinfo {author} {\bibfnamefont
  {T.}~\bibnamefont {Rabczuk}},\ }\href {\doibase 10.1016/j.cma.2019.112790}
  {\bibfield  {journal} {\bibinfo  {journal} {Computer Methods in Applied
  Mechanics and Engineering}\ }\textbf {\bibinfo {volume} {362}},\ \bibinfo
  {pages} {112790} (\bibinfo {year} {2020})}\BibitemShut {NoStop}%
\bibitem [{\citenamefont {Ranade}\ \emph {et~al.}(2021)\citenamefont {Ranade},
  \citenamefont {Hill},\ and\ \citenamefont {Pathak}}]{Ranade2021}%
  \BibitemOpen
  \bibfield  {author} {\bibinfo {author} {\bibfnamefont {R.}~\bibnamefont
  {Ranade}}, \bibinfo {author} {\bibfnamefont {C.}~\bibnamefont {Hill}}, \ and\
  \bibinfo {author} {\bibfnamefont {J.}~\bibnamefont {Pathak}},\ }\href
  {\doibase 10.1016/j.cma.2021.113722} {\bibfield  {journal} {\bibinfo
  {journal} {Computer Methods in Applied Mechanics and Engineering}\ }\textbf
  {\bibinfo {volume} {378}},\ \bibinfo {pages} {113722} (\bibinfo {year}
  {2021})}\BibitemShut {NoStop}%
\bibitem [{\citenamefont {Raissi}\ and\ \citenamefont
  {Karniadakis}(2018)}]{Raissi2018}%
  \BibitemOpen
  \bibfield  {author} {\bibinfo {author} {\bibfnamefont {M.}~\bibnamefont
  {Raissi}}\ and\ \bibinfo {author} {\bibfnamefont {G.~E.}\ \bibnamefont
  {Karniadakis}},\ }\href {\doibase 10.1016/j.jcp.2017.11.039} {\bibfield
  {journal} {\bibinfo  {journal} {Journal of Computational Physics}\ }\textbf
  {\bibinfo {volume} {357}},\ \bibinfo {pages} {125} (\bibinfo {year}
  {2018})}\BibitemShut {NoStop}%
\bibitem [{\citenamefont {Regazzoni}\ \emph {et~al.}(2019)\citenamefont
  {Regazzoni}, \citenamefont {Ded{\`{e}}},\ and\ \citenamefont
  {Quarteroni}}]{Regazzoni2019}%
  \BibitemOpen
  \bibfield  {author} {\bibinfo {author} {\bibfnamefont {F.}~\bibnamefont
  {Regazzoni}}, \bibinfo {author} {\bibfnamefont {L.}~\bibnamefont
  {Ded{\`{e}}}}, \ and\ \bibinfo {author} {\bibfnamefont {A.}~\bibnamefont
  {Quarteroni}},\ }\href {\doibase 10.1016/j.jcp.2019.07.050} {\bibfield
  {journal} {\bibinfo  {journal} {Journal of Computational Physics}\ }\textbf
  {\bibinfo {volume} {397}},\ \bibinfo {pages} {108852} (\bibinfo {year}
  {2019})}\BibitemShut {NoStop}%
\bibitem [{\citenamefont {Albash}\ and\ \citenamefont
  {Lidar}(2018)}]{Albash2018}%
  \BibitemOpen
  \bibfield  {author} {\bibinfo {author} {\bibfnamefont {T.}~\bibnamefont
  {Albash}}\ and\ \bibinfo {author} {\bibfnamefont {D.~A.}\ \bibnamefont
  {Lidar}},\ }\href {\doibase 10.1103/revmodphys.90.015002} {\bibfield
  {journal} {\bibinfo  {journal} {Reviews of Modern Physics}\ }\textbf
  {\bibinfo {volume} {90}} (\bibinfo {year} {2018}),\
  10.1103/revmodphys.90.015002}\BibitemShut {NoStop}%
\bibitem [{\citenamefont {Kadowaki}\ and\ \citenamefont
  {Nishimori}(1998)}]{Kadowaki1998}%
  \BibitemOpen
  \bibfield  {author} {\bibinfo {author} {\bibfnamefont {T.}~\bibnamefont
  {Kadowaki}}\ and\ \bibinfo {author} {\bibfnamefont {H.}~\bibnamefont
  {Nishimori}},\ }\href {\doibase 10.1103/physreve.58.5355} {\bibfield
  {journal} {\bibinfo  {journal} {Physical Review E}\ }\textbf {\bibinfo
  {volume} {58}},\ \bibinfo {pages} {5355} (\bibinfo {year}
  {1998})}\BibitemShut {NoStop}%
\bibitem [{\citenamefont {Kato}(1950)}]{kato1950adiabatic}%
  \BibitemOpen
  \bibfield  {author} {\bibinfo {author} {\bibfnamefont {T.}~\bibnamefont
  {Kato}},\ }\href@noop {} {\bibfield  {journal} {\bibinfo  {journal} {Journal
  of the Physical Society of Japan}\ }\textbf {\bibinfo {volume} {5}},\
  \bibinfo {pages} {435} (\bibinfo {year} {1950})}\BibitemShut {NoStop}%
\bibitem [{\citenamefont {Messiah}(1962)}]{messiah1962quantum}%
  \BibitemOpen
  \bibfield  {author} {\bibinfo {author} {\bibfnamefont {A.}~\bibnamefont
  {Messiah}},\ }\href@noop {} {\enquote {\bibinfo {title} {Quantum mechanics,
  vol. ii},}\ } (\bibinfo {year} {1962})\BibitemShut {NoStop}%
\bibitem [{\citenamefont {Sarandy}\ \emph {et~al.}(2004)\citenamefont
  {Sarandy}, \citenamefont {Wu},\ and\ \citenamefont
  {Lidar}}]{sarandy2004consistency}%
  \BibitemOpen
  \bibfield  {author} {\bibinfo {author} {\bibfnamefont {M.~S.}\ \bibnamefont
  {Sarandy}}, \bibinfo {author} {\bibfnamefont {L.-A.}\ \bibnamefont {Wu}}, \
  and\ \bibinfo {author} {\bibfnamefont {D.~A.}\ \bibnamefont {Lidar}},\
  }\href@noop {} {\bibfield  {journal} {\bibinfo  {journal} {Quantum
  Information Processing}\ }\textbf {\bibinfo {volume} {3}},\ \bibinfo {pages}
  {331} (\bibinfo {year} {2004})}\BibitemShut {NoStop}%
\bibitem [{\citenamefont {D-Wave}(2021)}]{dwave_qpu_2021}%
  \BibitemOpen
  \bibfield  {author} {\bibinfo {author} {\bibnamefont {D-Wave}},\ }\href@noop
  {} {\enquote {\bibinfo {title} {Technical description of the d-wave quantum
  processing unit},}\ } (\bibinfo {year} {2021}),\ \Eprint
  {http://arxiv.org/abs/D-Wave User Manual 09-1109A-X} {D-Wave User Manual
  09-1109A-X} \BibitemShut {NoStop}%
\bibitem [{\citenamefont {Glover}\ and\ \citenamefont
  {Kochenberger}(2018)}]{glover2018tutorial}%
  \BibitemOpen
  \bibfield  {author} {\bibinfo {author} {\bibfnamefont {F.}~\bibnamefont
  {Glover}}\ and\ \bibinfo {author} {\bibfnamefont {G.}~\bibnamefont
  {Kochenberger}},\ }\href@noop {} {\bibfield  {journal} {\bibinfo  {journal}
  {arXiv preprint arXiv:1811.11538}\ } (\bibinfo {year} {2018})}\BibitemShut
  {NoStop}%
\bibitem [{\citenamefont {Rogers}\ and\ \citenamefont
  {Singleton}(2020)}]{Rogers2020}%
  \BibitemOpen
  \bibfield  {author} {\bibinfo {author} {\bibfnamefont {M.~L.}\ \bibnamefont
  {Rogers}}\ and\ \bibinfo {author} {\bibfnamefont {R.~L.}\ \bibnamefont
  {Singleton}},\ }\href {\doibase 10.3389/fphy.2020.00265} {\bibfield
  {journal} {\bibinfo  {journal} {Frontiers in Physics}\ }\textbf {\bibinfo
  {volume} {8}} (\bibinfo {year} {2020}),\ 10.3389/fphy.2020.00265}\BibitemShut
  {NoStop}%
\bibitem [{Note1()}]{Note1}%
  \BibitemOpen
  \bibinfo {note} {The development of the QBoost method applied to regression
  problems and application to solve this PDE, which recovers the Burgers's
  equation as particular case, was motivated by the Airbus Quantum Computing
  Challenge, www.air bus.com/qc-challenge.html.}\BibitemShut {Stop}%
\bibitem [{\citenamefont {Burgers}(1948)}]{BURGERS1948171}%
  \BibitemOpen
  \bibfield  {author} {\bibinfo {author} {\bibfnamefont {J.}~\bibnamefont
  {Burgers}},\ }\href {\doibase https://doi.org/10.1016/S0065-2156(08)70100-5}
  {\emph {\bibinfo {title} {A Mathematical Model Illustrating the Theory of
  Turbulence}}},\ edited by\ \bibinfo {editor} {\bibfnamefont {R.}~\bibnamefont
  {{Von Mises}}}\ and\ \bibinfo {editor} {\bibfnamefont {T.}~\bibnamefont {{Von
  Kármán}}},\ \bibinfo {series} {Advances in Applied Mechanics},
  Vol.~\bibinfo {volume} {1}\ (\bibinfo  {publisher} {Elsevier},\ \bibinfo
  {year} {1948})\ pp.\ \bibinfo {pages} {171--199}\BibitemShut {NoStop}%
\end{thebibliography}%

\appendix

\section{\label{sec:B} Regularization term}

\begin{figure}[H]
\centering
\includegraphics[width=0.5\columnwidth]{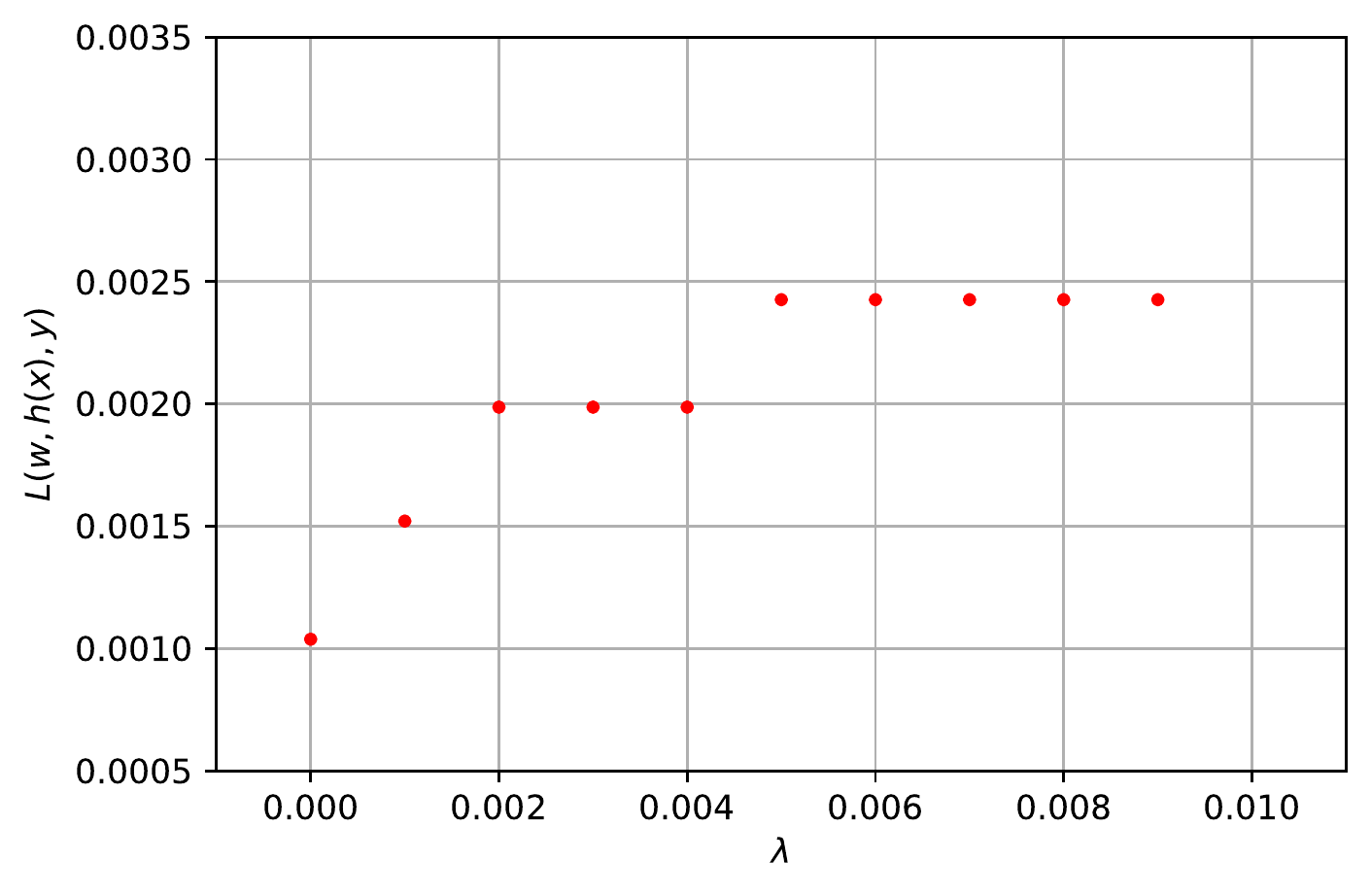}%
\caption{\label{fig:lossxl} Cost function, Eq. \ref{eq:loss}, as function of the regularization parameter, $\lambda$.}
\end{figure}

The dependence of the loss function on the regularization term is shown in Fig. \ref{fig:lossxl}. We see that that any increase in the regularization term results in an increase of the loss function, which is expected since it controls overfitting dislocating the objective function further from the global minimum. Therefore, we conclude that the best result occurs when the regularization term is null. Our conclusion is supported by the fact that we did not observe the presence of overfitting or underfitting in Fig. \ref{fig:ensemble}.

\end{document}